\documentclass{article}
\usepackage[auth-sc]{authblk}
\usepackage[toc,page]{appendix}

\usepackage{amssymb}
\usepackage{amsthm}
\usepackage{amsmath}
\usepackage{booktabs}
\usepackage{graphicx}
\usepackage[dvipsnames]{xcolor}
\usepackage{tabu}
\usepackage{here}
\RequirePackage[round]{natbib}

\usepackage{rotating}
\usepackage{fancyhdr}
\usepackage{etoolbox}
\usepackage{array,hhline}
\usepackage{multicol}
\usepackage{dsfont}
\usepackage{float}
\usepackage{amscd}
\usepackage{multirow}
\usepackage{colortbl}
\usepackage[margin=1in]{geometry}
\usepackage{setspace}
\usepackage{marginnote}
\usepackage{nameref}

\usepackage[utf8]{inputenc}

\RequirePackage[hypertexnames=false,bookmarks=true,colorlinks,citecolor=blue,urlcolor=blue]{hyperref}
\usepackage{bookmark}
\graphicspath{{./figures/}}

\newcommand{\eps}{\varepsilon}

\newcommand{\brho}{\rho}
\newcommand{\bphi}{\boldsymbol{\phi}}
\newcommand{\blam}{\boldsymbol{\lambda}}

\newcommand{\bEta}{\boldsymbol{\eta}}
\newcommand{\bth}{\boldsymbol{\theta}}
\newcommand{\blambda}{\boldsymbol{\lambda}}

\newcommand{\bH}{\boldsymbol{H}}
\newcommand{\bJ}{\boldsymbol{J}}

\newtheorem{proposition}{Proposition}

\newtheorem{theorem}[proposition]{Theorem}
\theoremstyle{remark}

\newtheorem{lemma}[proposition]{Lemma}

\begin{document}

\title{Robust estimation for Threshold  Autoregressive Moving-Average models}
\author[1]{Greta Goracci}
\author[1]{Davide Ferrari}
\author[2]{Simone Giannerini}
\author[1,3]{Francesco Ravazzolo}

\affil[1]{Faculty of Economics and Management, Free University of Bozen-Bolzano, Italy}
\affil[2]{Department of Statistical Sciences, University of Bologna, Italy}
\affil[3]{Department of Data Science and Analytics, BI Norwegian Business School, Norway}
\date{}
\maketitle

\begin{abstract}
Threshold autoregressive moving-average (TARMA) models are popular in time series analysis due to their ability to parsimoniously describe several complex dynamical features. However, neither theory nor estimation methods are currently available when the data present heavy tails or anomalous observations, which is often the case in applications. In this paper, we provide the first theoretical framework for robust M-estimation for TARMA models and also study its practical relevance. Under mild conditions, we show that the robust estimator for the threshold parameter is super-consistent, while the estimators for autoregressive and moving-average parameters are strongly consistent and asymptotically normal. The Monte Carlo study shows that the M-estimator is superior, in terms of both bias and variance, to the least squares estimator, which can be heavily affected by outliers. The findings suggest that robust M-estimation should be generally preferred to the least squares method. Finally, we apply our methodology to a set of commodity price time series; the robust TARMA fit presents smaller standard errors and leads to superior forecasting accuracy compared to the least squares fit. The results support the hypothesis of a two-regime, asymmetric nonlinearity around zero, characterised by slow expansions and fast contractions.
\end{abstract}
\noindent%
\section{Introduction}\label{sec:intro}
%
Threshold models are popular tools used to describe complex phenomena in many fields, including economics, finance, ecology, epidemiology \citep{Ton90,Cha09, giordani2007unified, Ton11,Han11, Cha17b}. Non-linearity is introduced by a  thresholding mechanism which implies multiple linear regimes; this enables the description of complex non-linear dynamical features, such as jumps, limit cycles, time irreversibility, while retaining good interpretability. Since their introduction by \cite{Ton78}, threshold  models have been widely studied, especially in their autoregressive specification, the so-called threshold autoregressive (TAR) models. In analogy with autoregressive moving-average (MA) models, threshold autoregressive moving-average (TARMA) models extend TAR models by including moving-average components in each regime \citep{Ton17}.
\par

Although technically more challenging than  TAR  models due to their non-Markovian nature,  TARMA models provide a powerful yet simple framework for many research problems involving non-linear phenomena; e.g., see \cite{Ton17}, \cite{Gor20} and \cite{Gor21}.  Nonetheless, their theoretical development has halted for many years and only recently \citet{Cha19} solved the long-standing open problem regarding the probabilistic structure of the first order TARMA model.  TARMA models possess a number of desirable features, including the following: they include moving-average (MA) components within a parametric non-linear setting;   they naturally account for measurement errors;  they are able to describe a wide range of long-run probabilistic behaviors spanning from transience to ergodicity, and even geometric ergodicity.  Also, the threshold framework  provides a natural way to describe series that appear to behave like random walks, when this behavior is incompatible with the theory underlying the data generating process.  One example in economics is the well-known purchasing power parity puzzle, which has motivated the development of unit-root tests where the alternative hypothesis is a stationary threshold model with a local unit-root regime \citep{End98,Bec04,Kap06,Bec08,Cha20b}. Tests for TARMA nonlinearity have been developed in \citet{Li11b} and \citet{Gor23}. \cite{Ang22} extend the results in \citet{Gor23} and develop a test for non-linear effects in the conditional mean for series with conditional heteroscedasticity by incorporating a GARCH specification.

\par

Compared to the simpler TAR models, estimation for the TARMA models is challenging due to the lack of a linear parameterization conditional on the threshold parameter.  Current inference methods mostly rely on the least squares (LS) approach \citep{Li11a}, which is knowingly influenced by outliers and heavy tails. Although aberrant observations are ubiquitous and appear in many  real applications \citep[e.g. see][]{giordani2007unified}, the issue of robust estimation and outlier detection for TARMA models has yet to be addressed from either theoretical or methodological viewpoints. On the other hand, robust estimation for the linear ARMA model has been extensively studied \citep[see][Chapter 8, for an overview]{maronna2019robust}. For the special case of the TAR model, \cite{chan1994robust} study the effect of additive outliers in LS estimation and propose a generalized M-estimation to mitigate the severe bias of the estimates. \cite{zhang2009note} consider a general class of robust estimators for threshold autoregressive models  and show consistency under   regularity conditions. \cite{grossi2015robust} compare  the relative efficiency of M-estimators
for the TAR  models to that of the LS estimator, showing that the former perform well when the error follows  heavy tailed or non-Gaussian distributions. \cite{van1999smooth} derive a robust estimation method for the parameters in smooth threshold TAR models, using generalized maximum likelihood estimation.
\par
Motivated by such a gap in the literature, we consider robust inference for the TARMA model using an  M-estimation approach. Our approach consists in replacing  the residual sum of squares criterion of \cite{Li11a} by a function with bounded derivative. This is a crucial feature which is necessary to gain stability of the estimates in the presence of different types of outliers. The resulting estimator for the autoregressive and moving-average parameters is shown to be strongly consistent and asymptotically normal, under standard regularity conditions. To the best of our knowledge, we provide the first results for robust estimation in parametric non-linear time series models with moving-average components.
\par
Similarly to the least squares estimator, the threshold parameter is found to be super-consistent with convergence rate of $n^{-1}$, while the autoregressive and moving-average parameters are root-$n$ consistent and asymptotically normal. The methodology is implemented using the special family of objective functions considered in \cite{Fer12} and \cite{la2015robust}, which include the LS estimator as a special case. While common contamination types are shown to increase, sometimes dramatically, bias and variance of the least squares estimator, our estimator mitigates the effect of observations that are incompatible with the assumed model, thus reducing the overall mean squared error.
\par
We showcase the performance of our new methodology by analyzing a set of  commodity price time series. Commodities are important in economics and finance due to their ability to anticipate the behavior of other macroeconomic variables; e.g., see \cite{Hamilton2011} and \cite{RR2013}. The predictive relationship between commodities and a number of macroeconomic variables is often non-linear, with asymmetric behavior depending on whether prices increase or decrease, e.g. see \cite{KV2011} and \cite{KV2013}.  Although TARMA models appear suitable for such data,  heavy tails, outliers and non-Gaussian innovations make the least squares estimate untrustworthy. The robust TARMA estimates generally provide a better fit with smaller standard errors  compared to non-robust estimates. Our robust TARMA specification confirms the existence of two dynamical regimes, separated by the threshold invariably located at zero, and corresponding to a slow, persistent growth (upper regime) and fast contractions (lower regime). The superior predictive performance of the robust TARMA approach can result in a key advancement in modelling the commodity market.
 \par
The remainder of the paper is organized as follows. In Section~\ref{sec:methods}, we describe the general M-estimation  approach. In Section~\ref{sec:properties}, we study the asymptotic behavior and robustness properties of the new estimator. In Section~\ref{sec:mc_study}, we study the finite-sample behavior of the new estimator and compare its robustness to the standard LS approach under common contamination models. In Section~\ref{sec:appl}, we apply the new method for robust estimation and outlier detection for several commodity price series. Conclusions and possible extensions of this work are presented in Section~\ref{sec:concl}. Further results from the analysis of commodity time series and technical proofs are reported in the Supplementary Material.
\section{Methodology}\label{sec:methods}

\subsection{Model setup and notation}\label{sec:notation}

Let  $\{X_t \}_{t \in \mathds{Z}}$ be the TARMA process defined by the difference equation
\begin{align}\label{eqn:TARMA}
X_t&=\begin{cases}
    \phi_{1,0}+\sum_{i=1}^p\phi_{1,i}X_{t-i}+\eps_t+\sum_{j=1}^{q} \theta_{1,j}\eps_{t-j}, & \mbox{if } X_{t-d}\leq r, \\
    \phi_{2,0}+\sum_{i=1}^p\phi_{2,i}X_{t-i}+\eps_t+\sum_{j=1}^{q} \theta_{2,j}\eps_{t-j}, & \mbox{if } X_{t-d}> r,
  \end{cases}
\end{align}
\noindent
where: $p \in \mathds{N}$ and $q\in \mathds{N}$ are, respectively, the autoregressive and  moving-average orders; the $\phi$'s and $\theta$'s are the autoregressive and moving-average parameters, respectively;   $1\leq d\leq D_0$, where $D_0\in \mathcal{D} \subset \mathds{N}$ is the delay parameter; $r\in \mathcal{R} \subseteq  \mathds{R}$ is the threshold parameter; and $\{\eps_{t}\}$ is the innovation process, with $E(\eps_t)=0$ and $E(\eps^2_t)=\sigma^2 <\infty$, which is usually assumed to be Gaussian white noise. Note that the TARMA model reduces to a linear ARMA as $|r|\rightarrow \infty$.

Equation~(\ref{eqn:TARMA}) defines two regimes, which will be referred to as lower and upper regimes  corresponding to $X_{r-d}\le r$ and $X_{r-d}> r$,  respectively. For each regime we have specific  parameter vectors defined as $
 \boldsymbol{\phi}_1=(\phi_{1,0},\dots,\phi_{1,p})^\intercal$, $\boldsymbol\theta_1=(\theta_{1,1},\ldots,\theta_{1,q})^\intercal$, $\boldsymbol{\phi}_2=(\phi_{2,0},\dots,\phi_{2,p})^\intercal$, $\boldsymbol\theta_2=(\theta_{2,1},\ldots,\theta_{2,q})^\intercal$, while
  $\bphi=(\bphi_1^\intercal,\bphi_2^\intercal)^\intercal$ and $\bth=(\bth_1^\intercal,\bth_2^\intercal)^\intercal$ are used to denote autoregressive and moving-average  parameters.
 The vector collecting all autoregressive and moving-average parameters is denoted by $\boldsymbol{\lambda}=(\boldsymbol{\phi}^\intercal_1,\boldsymbol{\phi}^\intercal_2, \boldsymbol{\theta}^\intercal_1,\boldsymbol{\theta}^\intercal_2)^\intercal\in\mathcal{Q}\subseteq\mathds{R}^{2(1+p+q)}$, while the overall parameter vector including also the threshold parameter is denoted by $\boldsymbol{\eta}=(\boldsymbol{\lambda}^\intercal,r,d)\in\mathcal{Q}\times\mathcal{R} \times \mathcal{D} :=\mathcal{H}$. We assume the parameter space $\mathcal{H}$ to be compact and  equipped with product metric. For simplicity of exposition, in this work we focus on TARMA models with a full model structure containing all lags up to order $p$ and $q$ for both regimes. However, our methodology can be applied without loss of generality to more complex order structures, including missing lags and specific orders for the upper and lower regimes.
\par

In many real-world applications, the process $X_t$  does not follow exactly the model specified in Equation~(\ref{eqn:TARMA}). Although the majority of the observations may be compatible with such model assumptions,  real data may diverge substantially from the assumed process due to the presence of heavy-tailed or asymmetric errors and aberrant observations. There are several models that may be used to represent the contamination process in the time series context.  One common model is the additive outlier (AO) model, which defines the contaminated process $X_t^\epsilon$ according to
$ X_t^\epsilon = X_t + Z^{\epsilon}_{t} W_t$,  where $X_t$ is the TARMA process defined in (\ref{eqn:TARMA}); $W_t$ is the contaminating process,  independent of $X_t$; and $Z^\epsilon_t$ is a binary process where $P(Z_t^{\epsilon} = 1) = \epsilon$ such that $\epsilon$ is the contamination level. Another common model is the replacement outlier (RO) model, where $X_t^\epsilon = (1-Z^{\epsilon}_{t})X_t + Z^{\epsilon}_{t} W_t$, with $W_t$, $X_t$ and $Z^\epsilon_t$  defined above. Finally, in the innovation outlier (IO) model the outliers affect not only the current observation, but also subsequent observations. IOs are obtained when the innovation $\eps_t$ follows a process different from the assumed nominal model. For example, $\eps_t$ is assumed to follow a Gaussian white noise process while the actual innovation process has the normal mixture distribution
$
(1-\epsilon)N(0, \sigma_0^2) + \epsilon N(0, \sigma_1^2)
$,
with $\sigma_0^2 \ll \sigma_1^2$. Outliers may also differ in their temporal structure. For example, patchy outliers arise from the AO and RO models by letting $Z_t$ be a Markov process remaining in one state for multiple time periods of fixed or random duration.

\subsection{Robust estimation}
For the time series $\{X_1,\dots,X_n\}$, define the residual function $\eps_t(\bEta) = X_t-E_{\boldsymbol{\eta}}[X_t|\mathcal{F}_{t-1}]$, $t=1,\dots, n$, where  $\mathcal{F}_{t}$ is the sigma algebra generated by $\{X_{t}, X_{t-1},\dots\}$. $E_{\boldsymbol{\eta}}[X_t|\mathcal{F}_{t-1}]$ denotes the expectation of $X_t$ conditional on the process history up to time $t-1$ and is computed with respect to the TARMA model described (\ref{eqn:TARMA}) with parameter $\bEta$. From (\ref{eqn:TARMA}), we have
\begin{align} \label{eq:res}
\eps_{t}(\boldsymbol{\eta})&=X_t-\left\{\phi_{1,0}+\sum_{i=1}^p \phi_{1,i}X_{t-i}+ \sum_{j=1}^{q}\theta_{1,j}\eps_{t-j}(\boldsymbol{\eta})\right\}I(X_{t-d}\leq r)\nonumber\\
&\phantom{=X_t}\,-\left\{\phi_{2,0}+\sum_{i=1}^p\phi_{2,i}X_{t-i}+ \sum_{j=1}^{q}\theta_{2,j}\eps_{t-j}(\boldsymbol{\eta})\right\}I(X_{t-d}> r).
\end{align}
An M-estimate $\hat \bEta_n$ of the parameter vector $\bEta$ is found by minimizing the objective function
\begin{equation}\label{eqn:rho}
  {\rho}_n(\boldsymbol{\eta})= \sum_{t=1}^{n}\rho\left(\dfrac{\eps_t(\bEta)}{\hat \sigma}\right),
\end{equation}
\noindent
where $\rho:\mathds{R} \mapsto \mathds{R}$ is a loss function  often referred to as  $\rho$-function in the literature of robust statistics, and $\hat \sigma$ is a robust estimate of scale which is obtained simultaneously with $\boldsymbol{\eta}$ as an $M$-scale estimate.  To obtain  robustness of $\hat \bEta_n$, we require the following standard conditions on $\rho$: (i) $\rho(z)$ is non-decreasing function of $|z|$; (ii) $\rho(0)=0$; (iii) $\rho(z)$ is increasing for $z>0$ such that $\rho(z)<\rho(\infty)$; and (iv) the derivative $\psi(z) = \partial \rho(z)/\partial z$ satisfies $|\psi(z)|<c$ for some finite constant $c>0$ and all $z \in \mathds{R}$. There  is a number of functions  satisfying the above requirements.

Here we study the $\rho$ function considered in \citet{Fer12} by taking $\rho(z) = -(f(z)^\alpha - 1)/\alpha$ for $\alpha>0$,  and $\rho(z)= - \log(f(z))$ for $\alpha=0$, where $f$ is the assumed probability density function for the innovations. For the special case of Gaussian innovations,  the objective function can be written as
 \begin{equation}\label{eqn:rhoF}
 \rho_n(\boldsymbol{\eta})= -
  \dfrac{1}{\alpha}\sum_{t=1}^{n} \left[ \left(2\pi \hat \sigma^2\right)^{-\alpha/2} \exp\left\{-\alpha\dfrac{\eps_t^2(\bEta)}{2\hat \sigma^2 } \right\}  - 1\right],
\end{equation}
for  $\alpha>0$. The limit case $\alpha \rightarrow 0$ corresponds to the maximum likelihood objective with
\begin{equation}\label{eqn:rhoF_ml}
\rho_n(\boldsymbol{\eta}) = - \dfrac{n}{2}\log(2\pi) - \dfrac{n}{2}\log(\hat \sigma^2)  +  \sum_{t=1}^{n} \dfrac{\eps_t^2(\bEta)}{2\hat \sigma^2}.
\end{equation}
When $\sigma^2$ is taken as known, minimizing (\ref{eqn:rhoF_ml}) is equivalent to minimizing the residual sum of squares $\sum_{t=1}^{n} \eps_t^2(\bEta)$. In this respect, the function of Equation~(\ref{eqn:rhoF}) represents a robust generalization of the well-established LS estimator for the TARMA model of \cite{Li11a}. The tuning parameter $\alpha$ controls the trade-off between efficiency and robustness of the underlying estimator; this makes this example particularly useful for analyzing the properties of the estimator for various degrees of robustness. For $\alpha>0$, the derivative
$
\psi(z) = \partial \rho(z)/\partial z = f^\alpha(z)  \partial \log f(z)/\partial z
$
is bounded for common family of density functions and $\psi(z) \rightarrow 0$ as $|z| \rightarrow \infty$, and corresponds to a re-descending M-estimator. On the other hand, for the limit case $\alpha \rightarrow 0$, we have $\rho(z) \rightarrow \log(f(z))$ and $\psi(z) = \partial \log f(z)/\partial z$. This case corresponds to the maximum likelihood estimator and the derivative $\psi(z)$ is typically unbounded, which leads to estimators that are sensitive to the presence of outliers.
\par
One practical hurdle in the derivation of $\hat \bEta_n$ is the discontinuity of ${\rho}_n(\boldsymbol{\eta})$ in $r$. To cope with this issue, the minimization is  carried out in two steps. First, given $r$ and $d$, we take the profile estimator of $\blambda$
\begin{equation}\label{eq:lambda_est}
\hat{\boldsymbol{\lambda}}_n(r, d)=\underset{{\boldsymbol{\lambda}\in \mathcal{Q}}}{\text{argmin}} \ {\rho}_n(\boldsymbol{\lambda},r, d),
\end{equation}
\noindent
and define ${\rho}^*_n(r)={\rho}_n(\hat{\boldsymbol{\lambda}}_n(r),r, d)$. Second, since ${\rho}^*_n(r, d)$ can only take a finite number of values, it can be minimized by searching over some grid $\widetilde{\mathcal{R}} \times \mathcal{D}$, i.e.,
$$(\hat{r}_n, \hat{d}_n)= \underset{(r,d) \in \widetilde{\mathcal{R}} \times \mathcal{D} }{\text{argmin}} \ {\rho}^*_n(r,d),
$$
where $\widetilde{\mathcal{R}}$  may be data-dependent. The final estimator is obtained by the plug-in method as
$$\hat{\boldsymbol{\eta}}_n=(\hat{\boldsymbol{\lambda}}^\intercal_n(\hat{r}_n, \hat{d}_n),\hat{r}_n, \hat{d}_n)^\intercal :=(\hat{\boldsymbol{\lambda}}^\intercal_n,\hat{r}_n, \hat{d}_n)^\intercal.$$
\noindent
Solving the minimization problem in (\ref{eq:lambda_est}) is  equivalent to finding the zeros of the weighted least squares estimating equations
\begin{align}\label{eqn:est_eqn}
\dfrac{\partial}{\partial \blambda} \rho_n(\bEta) =  \sum_{t=1}^n w(\eps_t(\bEta)) \dfrac{\partial \eps^2_t(\bEta)}{\partial \blambda} = \mathbf0,
\end{align}
where weights take the form $w(\eps_t(\bEta)) := \psi(|\eps_t(\bEta)|^{1/2})$. To ensure robustness, such weights must be relatively small when the residual is incompatible with the assumed distribution for the innovations, such as the Gaussian distribution. Solving directly (\ref{eqn:est_eqn}) in $\blambda$ may be computationally  difficult due to the presence of multiple local minima. This is typically the case for re-descending estimators for which the  derivative $\psi(u) = \rho'(u)$ is not monotone.
\par
To solve the above computational issues, we propose an iteratively re-weighted least squares (IRLS) approach to compute the estimates. The IRLS algorithm alternates two steps until convergence: (i) computing the weights $\tilde w_t = w(\eps_t(\tilde\bEta))$ using the current parameter value, say $\tilde \bEta$, and (ii)  updating the parameters by solving
$
\sum_{t=1}^n \tilde w_t \partial \eps^2_t(\bEta)/\partial \blambda = \boldsymbol{0}
$, which is equivalent to minimizing the weighted residual sum of squares
$
\sum_{t=1}^n \tilde w_t \eps^2_t(\bEta)
$. Note that, for fixed $r$, the parameter update from Step (ii) is just a weighted least squares problem which can be solved efficiently using existing algorithms for TARMA estimation.
\par
The above  IRLS approach is fast in execution, typically requiring only a few iterations to converge.
In all our numerical applications we use the following approach to obtain the initial estimate for the IRLS algorithm. We begin by trimming a percentage of the data corresponding to the most extreme observations; here we choose 10\%. Then we run the LS estimator on the trimmed sample. This allows us to obtain a fairly robust initial estimate not affecting the convergence properties of the algorithm.
\par
Standard errors for $\hat \blambda_n $ are computed using the asymptotic distribution of the estimator derived in Section \ref{sec:properties}. Particularly,   $\sqrt{n}\hat \blambda_n$   converges in distribution to a multivariate normal distribution with zero mean and covariance matrix  $\bH(\bEta)^{-1} \bJ(\bEta) \bH(\bEta)^{-1}$, where  $\bH(\bEta)$ and $\bJ(\bEta)$ are, respectively, the sensitivity and variability matrices whose expression is given in Theorem~\ref{thm:rate}. The asymptotic variance can be estimated consistently using the sandwich estimator $\hat \bH(\hat \bEta_n)^{-1} \hat \bJ(\hat \bEta_n) \hat \bH(\hat \bEta_n)^{-1}$ where
\begin{align} \label{eqn:sens_var}
\hat{\bH}(\bEta)  = \dfrac{1}{n} \sum_{t=1}^n \frac{\partial^2\rho(\eps_t(\bEta))}{\partial\blam\partial\blam^\intercal} ,  \ \ \hat{\bJ}(\bEta)  = \dfrac{1}{n}\sum_{t=1}^n \left(\frac{\partial\rho(\eps_t(\bEta))}{\partial\blam}\right) \left(\frac{\partial\rho(\eps_t(\bEta))}{\partial\blam}\right)^\intercal
\end{align}
are estimates of the sensitivity and variability matrices  $\bH(\bEta)$ and $\bJ(\bEta)$.
\section{Large sample properties}\label{sec:properties}

In this section, we study the behavior for the estimator $\hat \bEta_n$ as $n$ diverges. We use $\bEta_0$ to denote the minimizer of the population objective
\begin{equation}\label{eqn:objective_pos}
\rho^\dag(\bEta) =  E \left[ \rho( \eps_t(\bEta)\right].
\end{equation}
In the rest of this section, we assume that $\bEta_0$ exists and is unique. Note that here, differently from previous works on TARMA estimation, the true process generating the data does not necessarily coincide with the nominal TARMA model described in Section \ref{sec:notation} and the expectation in (\ref{eqn:objective_pos}) may be taken with respect to a process outside the TARMA model family. In this case,  the population parameter $\bEta_0$ should be regarded as the optimal process in terms of minimizing the density divergence implied by $\rho$ between the parametric TARMA model and the actual process underlying the data.

For the results presented in the remainder of this section, we require the following regularity conditions:
\begin{description}
  \item[(A1)] $\{X_t\}$ is invertible, strictly stationary and ergodic.
  \item[(A2)] $\{\eps_t\}$ has  bounded continuous and positive density on the real line; moreover, $E[\eps_t]=0$ for each $t$.
  \item[(A3)] $\rho(\cdot)$ is non-decreasing and has bounded first derivative $\rho'(\cdot)$ such that $\rho'(x)=0$ if and only if $x=0$. Moreover, the function $l(x)=E[\rho(\eps_t+x)-\rho(\eps_t)]$, $x\in\mathds{R}$, is continuous at 0 and $l(x)=0$ if $x=0$ whereas $l(x)>0$ if $x\neq 0$.
  \item[(A4)] There exist non-random vectors $\mathbf{a}=(1,a_1,\dots,a_p)^\intercal\in \mathds{R}^{p+1}$, with $a_d=r_0$ and $\mathbf{b}\in\mathds{R}^q$ such that
      $(\bphi_{10}-\bphi_{20})^\intercal\mathbf{a}+(\bth_{10}-\bth_{20})^\intercal\mathbf{b}\neq0$.
  \item[(A5)] $\rho(\cdot)$ has bounded second derivative $\rho''(\cdot)$.
\end{description}
Assumptions (A1), (A2) and (A4) are standard requirements in the threshold framework. Regarding Assumption (A1), more details on the conditions ensuring stationarity and ergodicity of TARMA models are given by \citet{Lin99} and \citet{Cha19}, while invertibility is studied in \citet{Cha10}. A discussion on the invertibility of threshold moving-average models can also be found in \citet{Lin05} and \citet{Lin07}. Assumption (A3) is a basic requirement  for robustness. For instance, the re-descending estimator in \cite{Fer12} satisfies these properties for common families of distributions for the innovation process. Another possible choice for $\rho$ leading to similar robustness properties is Tukey's bisquare function \citep[e.g., see]{maronna2019robust}.  Assumption (A4) is the same condition considered in \cite{Li11a} in order to ensure threshold identification.    Assumption (A5) is stronger than Assumption (A3), and is needed to guarantee a regular behavior of the expansion leading to asymptotic normality for the ARMA parameter estimator in the two regimes.
\par
The loss function $\rho$ should satisfy at least the Fisher consistency property. Namely, when the data are generated by a TARMA process with parameter $\bEta_0$, then $\bEta_0$ should be also the minimizer of the population objective $\rho^\dag(\bEta) = E_{\bEta_0}[\rho(\eps_t(\bEta))]$, where expectation is taken with respect to the TARMA process with parameter $\bEta_0$. Following steps analogous to Lemma 1 in \cite{Fer12}, one can show that the re-descending estimator minimizing (\ref{eqn:rhoF}) is Fisher consistent for the parameter $\bEta$ for any $\alpha > 0$. The special case $\alpha=0$ corresponds to the maximum likelihood estimator, which is clearly Fisher consistent, but  does not satisfy Assumption (A3) and leads to estimates that are influenced by outliers.
\par
 The next theorem shows the strong consistency of the estimator $\hat{\bEta}_n$.
\begin{theorem}\label{thm:cons}
  Under Assumptions (A1) -- (A3) and (i) $E[X_t^2]<\infty$, (ii) $E[\eps_t^2]<\infty$ and (iii) $\bphi_{10}\neq\bphi_{20}$ or $\bth_{10}\neq\bth_{20}$, we have
$\hat{\boldsymbol{\eta}}_n\xrightarrow[n\to\infty]{a.s.}\boldsymbol{\eta}_0$.
\end{theorem}
In the following, we derive the convergence rates of $\hat{r}_n$ and $\hat{\boldsymbol{\lambda}}_n$ and  prove the uniform asymptotic normality of $\hat{\blam}_n$.  To this end, let $\partial\rho(\bEta_0)/\partial\blam$ and $\partial^2\rho(\bEta_0)/\partial\blam\partial\blam^\intercal$ be the first and the second derivative of the function $\rho(\bEta)$ with respect to $\blam$ evaluated at the parameter vector $\bEta_0$. Moreover, define
\begin{align}\label{eqn:matrices_def}
   \boldsymbol{H}(\bEta)  = E\left[\frac{\partial^2\rho(\bEta)}{\partial\blam\partial\blam^\intercal}\right], \ \ \boldsymbol{J}(\bEta) =   E\left[\left(\frac{\partial\rho(\bEta)}{\partial\blam}\right) \left(\frac{\partial\rho(\bEta)}{\partial\blam}\right)^\intercal\right].
  \end{align}
  The matrices $\boldsymbol{H}(\bEta)$ and $\boldsymbol{J}(\bEta)$, evaluated at $\bEta_0$ form the asymptotic variance $\boldsymbol{H}(\bEta_0)^{-1}\boldsymbol{J}(\bEta_0)\boldsymbol{H}(\bEta_0)^{-1}$  for $\hat{\blam}_n$. Hence, we require the following assumptions.
  \begin{description}
    \item[(A6)] The matrices $\boldsymbol{H}(\bEta_0)$ and $\boldsymbol{J}(\bEta_0)$ exist and $\boldsymbol{H}(\bEta_0)$ is invertible.
  \end{description}

\begin{theorem}\label{thm:rate}

  Under Assumptions (A1) - (A4) and (i) $E[X_t^4]<\infty$, (ii) $\{\eps_t\}$ is a sequence of i.i.d. random variables with $E[\eps_t^4]<\infty$ it holds that:
\begin{description}
  \item[(i)] $n(\hat{r}_n-r_0)=O_p(1)$.
  \end{description}
  Moreover, if also  Assumptions (A5) and (A6) hold, it follows that
  \begin{description}
  \item[(ii)] $\sqrt{n}\sup_{|r-r_0|\leq B/n}\left\|\hat{\boldsymbol{\lambda}}_n(r)-\hat{\boldsymbol{\lambda}}_n(r_0)\right\|=o_p(1),$ for any fixed constant $B>0$;
  \item[(iii)]
  $\sqrt{n}(\hat{\boldsymbol{\lambda}}_n-\boldsymbol{\lambda}_0)\xrightarrow[n\to\infty]{d}N(\boldsymbol0,\boldsymbol{H}(\bEta_0)^{-1} \boldsymbol{J}(\bEta_0), \boldsymbol{H}(\bEta_0)^{-1})$.
\end{description}
\end{theorem}

The proofs of Theorems~\ref{thm:cons} and \ref{thm:rate} follow an approach similar to  \cite{Kou03} and \cite{Li11a} with some notable differences. While   \cite{Kou03} also focus on general M-estimators, their results rely heavily on the simpler structure of the autoregressive process, while here we also take   into account  the moving-average component. \cite{Li11a} consider both autoregressive and moving-average components, but their proofs are only valid for the specific case of the least squares objective function, which is much simpler to handle than generic M-estimating functions.
\par
Finally, note that the estimator of the threshold $\hat r_n$ is super-consistent. In practical terms, this means that the threshold can be taken as given, provided the sample size is adequate. For this reason we have omitted the derivation of the robust asymptotic distribution for $\hat r_n$.
\section{Monte Carlo study}\label{sec:mc_study}

In this section, we perform a Monte Carlo study to assess the performance of our  robust estimator. We consider the four parameter settings  shown in Table  \ref{tab:true} for the following TARMA$(1,1)$ process:
\begin{align}\label{eqn:TARMA11}
X_t&=\begin{cases}
    \phi_{1,0}+\phi_{1,1}X_{t-1}+\eps_t+ \theta_{1,1}\eps_{t-1}, & \mbox{if } X_{t-d}\leq r, \\
    \phi_{2,0}+\phi_{2,1}X_{t-1}+\eps_t+ \theta_{2,1}\eps_{t-1}, & \mbox{if } X_{t-d}> r,
  \end{cases}
\end{align}
where $\eps_t \sim   N(0,1)$ and $d=1$. The choice of parameters reflects different long-run probabilistic behaviors of the TARMA process. In particular, Cases 1 and 3 correspond to ergodic processes, whereas Cases 2 and 4 correspond to   geometrically ergodic processes. Also,  Case 3 has unit roots in both regimes but is globally stationary; this is a challenging case laying on the boundary of the ergodicity region; see \cite{Cha19} for more details.

The data generated from the clean TARMA process are contaminated using a fraction $\epsilon$ of outliers. In particular, we consider both additive outliers (AOs) and innovation outliers (IOs) corresponding to the two Monte Carlo experiments described below.

\begin{table}
\caption{\label{tab:true} Parameter settings for the TARMA$(1,1)$ model in Equation~(\ref{eqn:TARMA11}). Cases 1 and 3 represent ergodic processes, whereas Cases 2 and 4 correspond to geometrically ergodic processes. Case 3 has unit roots in both regimes but is globally stationary.}
\centering
\begin{tabular}[t]{lrrrrrrrr}
  & $\phi_{1,0}$ & $\phi_{1,1}$ & $\theta_{1,1}$ & & $\phi_{2,0}$ & $\phi_{2,1}$ & $\theta_{2,1}$ & $r$\\
\cmidrule(lr){2-4}\cmidrule(lr){6-8}\cmidrule(lr){9-9}
Case 1 & 0.5 & -0.5 & -0.5 & &  0.0 & -1.0 &  0.5 & 0.2\\
Case 2 & 0.5 &  0.3 &  0.6 & &  1.0 & -0.5 & -0.4 & 0.2\\
Case 3 & 2.0 &  1.0 &  0.5 & & -1.5 &  1.0 & -0.5 & 0.2\\
Case 4 & 0.6 &  0.6 & -0.7 & & -1.0 &  0.4 &  0.5 & 0.2\\
\cmidrule(lr){2-4}\cmidrule(lr){6-8}\cmidrule(lr){9-9}
\end{tabular}
\end{table}

\begin{itemize}
 \item[]{\bf Monte Carlo Experiment 1: Additive outliers.} We consider the  family of contaminated processes
$X_t^{\epsilon, k} = (1-Z^{\epsilon}_{t})X_t + Z^{\epsilon}_{t} W^k_t$, where $X_t$ is the clean TARMA process in Eq.~(\ref{eqn:TARMA11});  $W_t^k = X_t + (-1)^{\xi_t} k$, with $k=10$, is the contaminating process; $\xi_t$ is a binary process such $P(\xi_t = 1) = 0.95$; and  $Z_t^{\epsilon}=1$  if $ t\mod \epsilon^{-1} = 0$ and $Z_t=0$ otherwise.
\item[] {\bf Monte Carlo Experiment 2: Innovation outliers.} The data are generated from the model in Eq.~(\ref{eqn:TARMA11}) by taking $\varepsilon_t \sim N(0,1) + k (-1)^\xi_t Z^\epsilon_{t}$, where $\xi_t$ is a binary process such that $P(\xi_t = 1) = 0.95$ , $Z_t^{\epsilon}=1$  if $ t\mod \epsilon^{-1} = 0$ and $Z_t=0$ otherwise.
\end{itemize}

To assess the performance of our robust methodology, we compute Monte Carlo estimates of the  bias,  $ \Vert E(\hat \bEta_n) - \bEta_0 \Vert^2_2,
$ and  variance, $ \Vert \text{var}(\hat \bEta_n)\Vert^2_2$ of our estimator, where $\bEta_0$ represents the parameter vector for the clean TARMA process and
$\Vert \cdot \Vert_2$ is the Euclidean norm. Estimates are based on 1000 Monte Carlo replications with sample size $n=100, 200$. In practice, in each contamination setting, we add 10\% of equally spaced outliers of size $k=10$ with random sign depending on $\xi_t$. Note that, differently from AOs,  IOs are much harder to treat since they enter the state equation and interact non-trivially with the non-linearity of the TARMA process. This can exert a long-term influence upon the series and even produce a qualitative change in the dynamics. The above experiments aim to mimic real scenarios encountered in economics and finance where  contamination may occur in both tails but are prevalent in one.
\par
\begin{figure}
  \centering
  \includegraphics[width=0.95\linewidth,keepaspectratio]{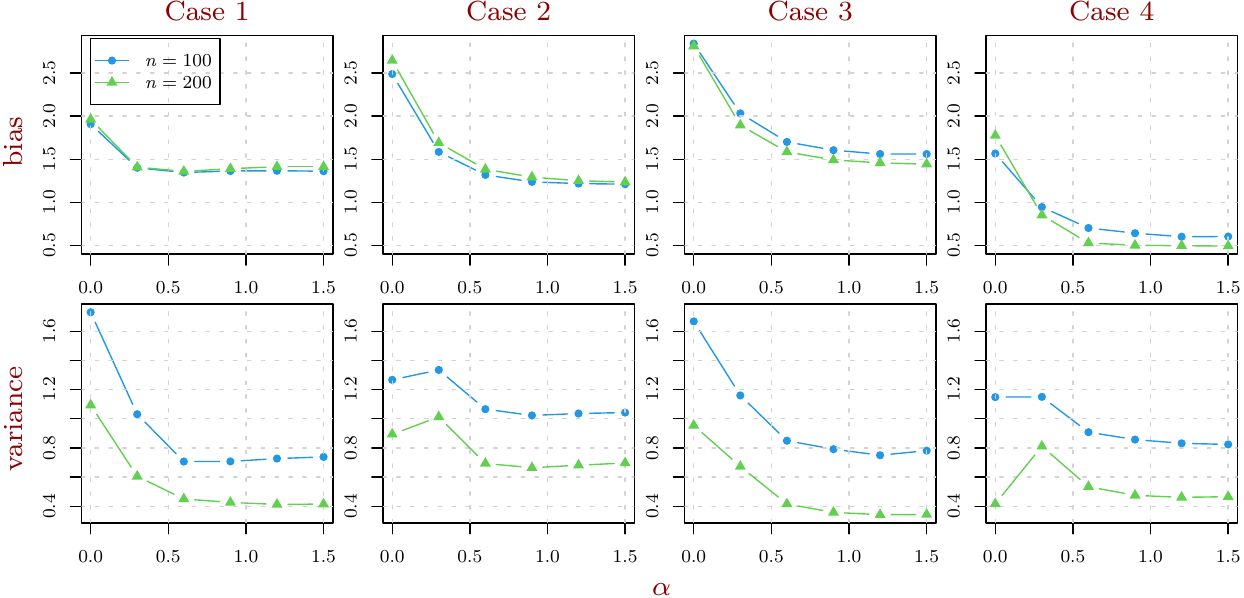}
  \caption{Results from Monte Carlo Experiment 1 (additive outliers). Bias and variance  as a function of the robustness parameter $\alpha$. Clean data are generated using the TARMA under the four parameterizations in Table~\ref{tab:true} (Cases 1--4), while contaminations are introduced by AOs.}\label{fig:MC1}
\end{figure}

Figure~\ref{fig:MC1} shows bias and variance for the four TARMA specifications under AO contamination for   values of the robustness parameter $\alpha = 0,0.3,0.6,0.9,1.2,1.5$, where $\alpha=0$ corresponds to the special case of the non-robust LS estimator. In all the  settings, the bias decreases significantly when $\alpha$ moves away from zero and stabilizes for values of $\alpha$ larger than $1$. Cases 2 and 4 show the sharpest decrease, which may be an effect due to the geometric ergodicity. Interestingly,  the variance also stabilizes starting for a value of $\alpha$ larger $1$; however, differently from the bias, here Cases 1 and 3 show the sharpest decrease. While both bias and variance of the LS estimator are considerably affected by outliers, the robust estimator with $\alpha \geq 1$ is generally successful in mitigating their influence.

\begin{figure}
  \centering
  \includegraphics[width=0.95\linewidth,keepaspectratio]{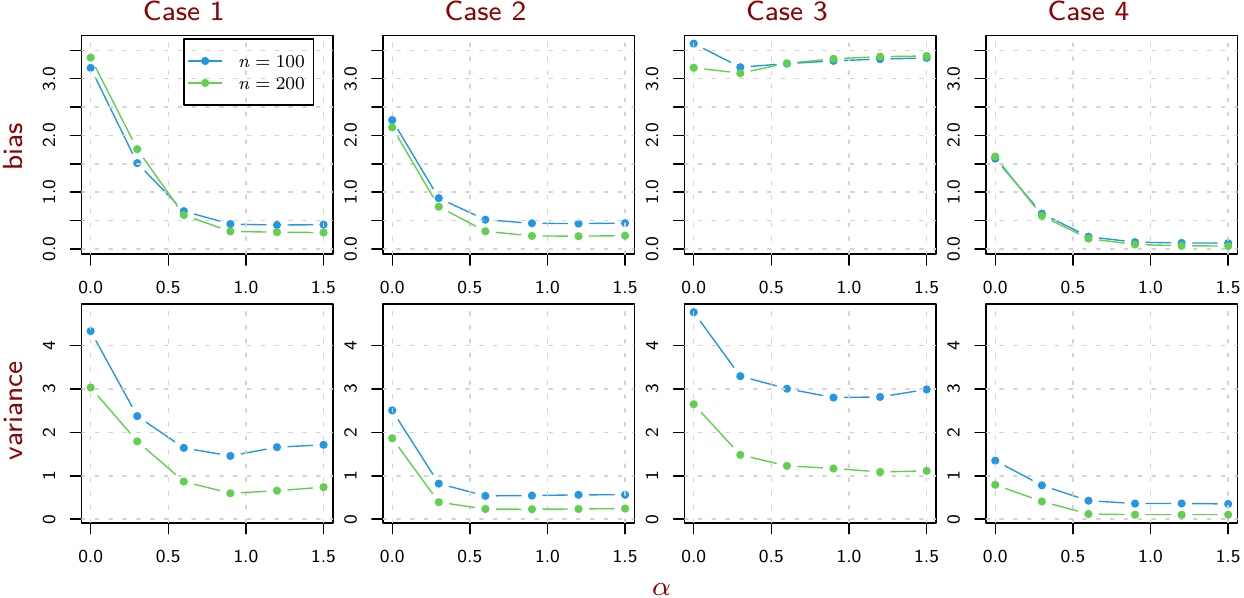}
  \caption{Results of Monte Carlo Experiment 2 (innovation outliers). Bias and variance   as a function of the robustness parameter $\alpha$. Clean data are generated using the TARMA under the four parameterizations in Table~\ref{tab:true} (Cases 1--4), while contaminations are introduced by IOs.}\label{fig:MC2}
\end{figure}

Figure \ref{fig:MC2} shows bias and variance for the four TARMA specifications under IO contamination.   The findings are consistent with those reported in Figure  \ref{fig:MC1}. In all the  scenarios, a value of $\alpha>0$ suffices to improve both bias and variance. The improvement is dramatic in most cases for sufficiently large $\alpha$. Note that the reduction in bias and variance is less marked for Case 3, which sits at the boundary of the parametric region of ergodicity. Even if the process is globally stationary, its regimes are both $I(1)$ so that the outlier effect decays very slowly.

\subsection{Asymptotic bias under contamination}\label{sec:bias}

The asymptotic bias under contamination is a common   measure of robustness for the time series framework. Other measures include the influence curve, introduced by \cite{hampel1974influence} in the i.i.d. framework, which measures the influence of infinitesimal outlier contamination on the parameter estimates. Also, \cite{martin1986influence} consider a generalization of influence functionals in time-series models based on a replacement outlier model. Here we focus on the asymptotic bias since it does not assume infinitesimal contaminations and provides a realistic representations of the behavior of the estimator in practical situations. Let $\hat \bEta_\infty(F)$ be the almost sure limit of the estimator $\hat \bEta_n = (\hat \blambda_n^\intercal, \hat r_n , \hat d_n)$ applied to a process with distribution $F$. The asymptotic squared bias for  $\hat \bEta_\infty$ applied to the contaminated process $\{X_t^{\epsilon, k}\}$ is given by
\begin{align*}
B(\hat \bEta_{\infty}, \bEta_0, \epsilon, k) = \left\Vert  \hat \bEta_{\infty}(F(X_t^{\epsilon, k})) - \bEta_0 \right\Vert^2_2,
\end{align*}
where $F(X_t^{\epsilon, k})$ is the distribution of the contaminated process $\{X_t^{\epsilon, k}\}$ and $\left\Vert \cdot \right\Vert_2$ is the Euclidean norm.
\par
\begin{figure}[htp]
  \centering
\includegraphics[width=0.7\linewidth,keepaspectratio]{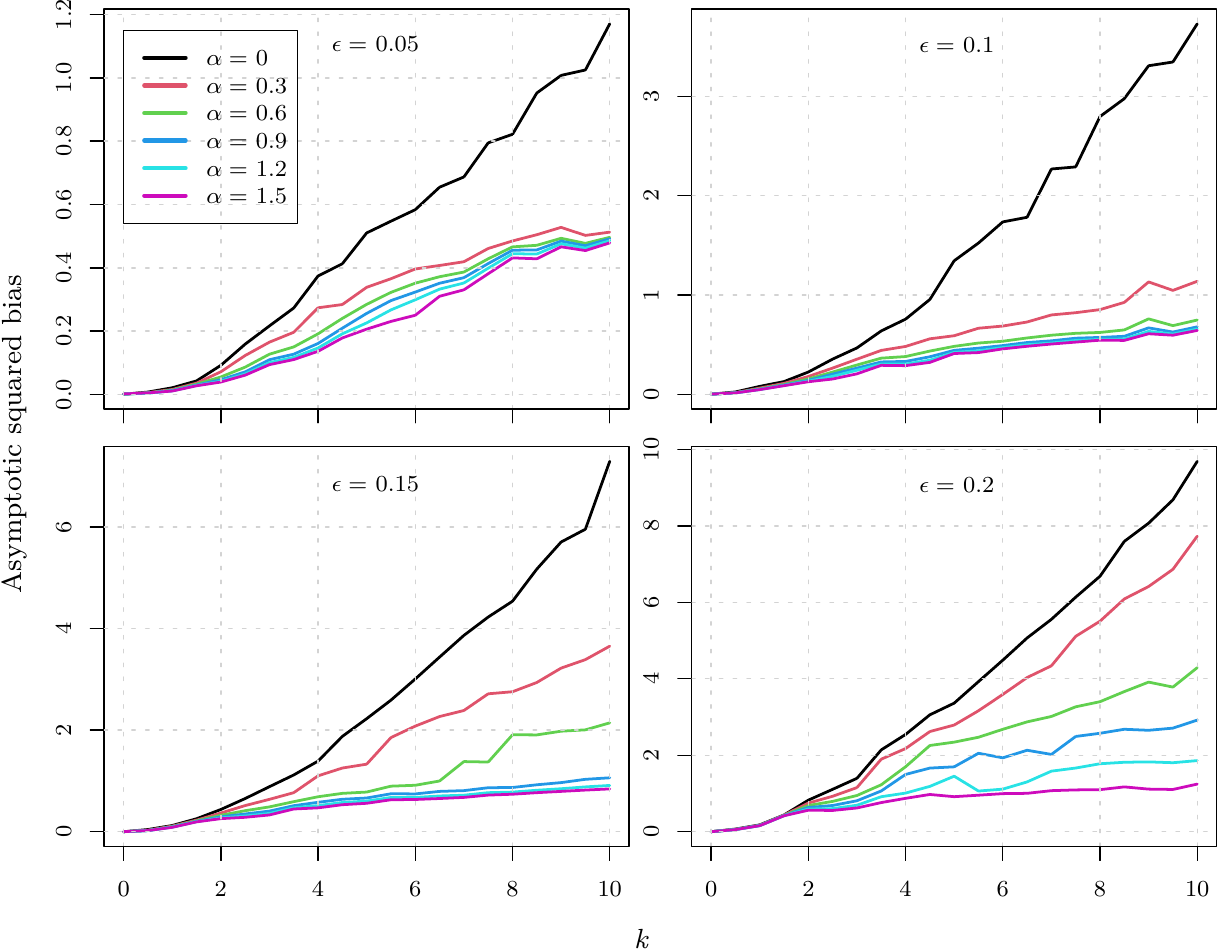}
  \caption{Asymptotic squared bias against outlier size $k$, under AO contamination, for contamination levels $\epsilon=0.05, 0.1, 0.15, 0.2$ and different values of the robustness parameter  $\alpha$. The contaminating AO process is described in MC Experiment 1.}\label{fig:MC3}
\end{figure}
\begin{figure}[htp]
  \centering
\includegraphics[width=0.7\linewidth,keepaspectratio]{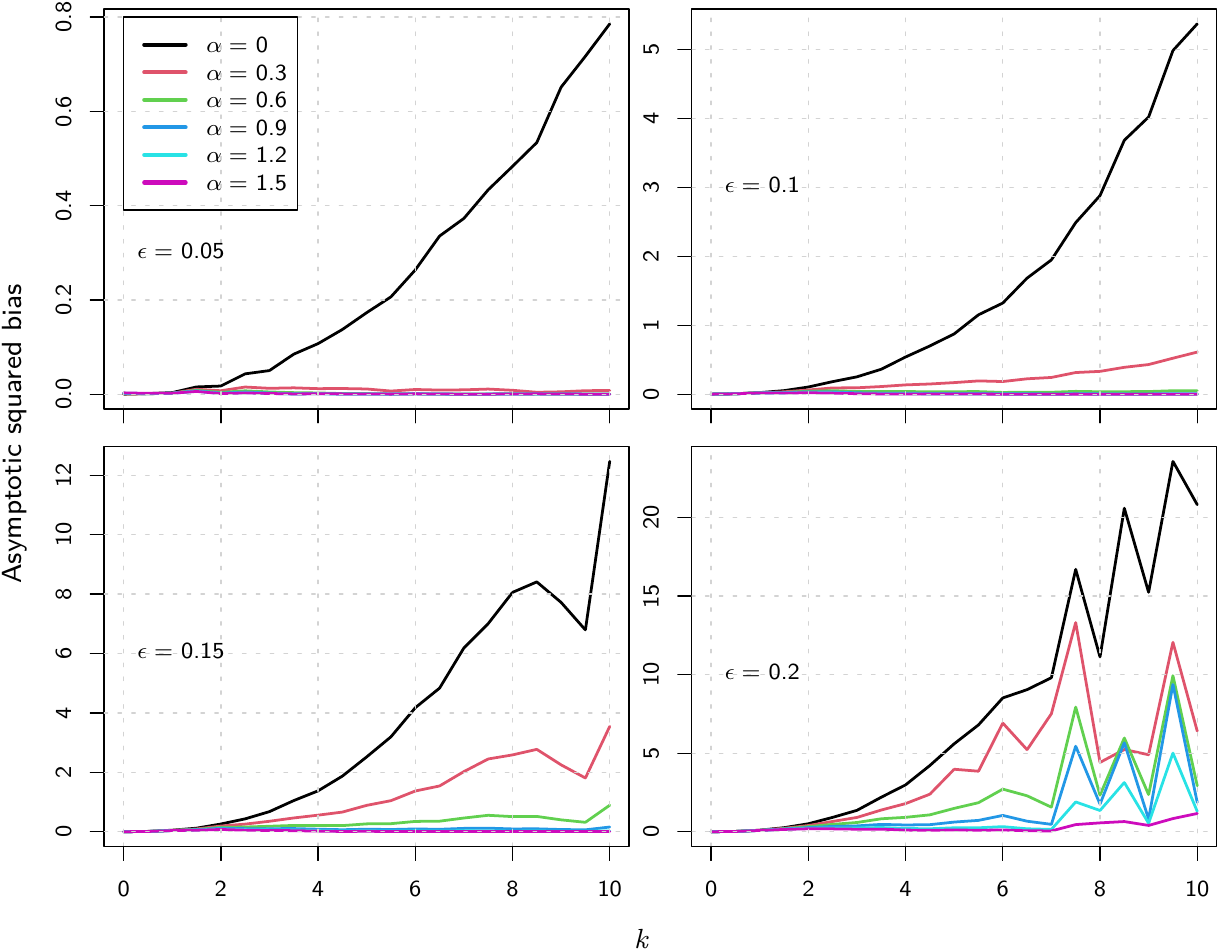}
  \caption{Asymptotic squared bias against outlier size $k$, under IO contamination, for contamination levels $\epsilon=0.05, 0.1, 0.15, 0.2$ and different values of the robustness parameter  $\alpha$. The contaminating IO process is described in MC Experiment 2.}\label{fig:MC4}
\end{figure}

In Figures~\ref{fig:MC3} and \ref{fig:MC4} we show the behavior of the asymptotic bias against outlier size $k$, for contamination levels $\epsilon=0.05, 0.1, 0.15, 0.2$ and different values for the robustness parameter $\alpha$. The asymptotic values are computed using series of size $n=20000$ and the plots summarize the four cases through the median. The behavior for both additive and innovation outliers is similar. The bias of the non-robust estimator ($\alpha=0$) diverges quickly as $k$ increases. On the other hand, for contamination levels up to 10\%, small values of $\alpha$ are enough to achieve robustness. As the contamination level increases, larger values of $\alpha$ are needed to stabilize the asymptotic bias. A value of $\alpha$ close to one achieves a remarkable robustness even when 20\% of the data are contaminated and in case of large outliers ($\epsilon=0.2$, lower right panels).

\section{Application: Analysis of commodity prices}\label{sec:appl}

Commodities are raw materials or primary agricultural products used as inputs in the production of other goods and are commonly traded in the cash market or as derivatives. Commodity prices are extremely important in individual, country-level economies: since they respond quickly to economic shocks, such as increase in demand, they are often used to predict the behavior of other economic variables.
\par
We consider $336$ monthly observations for the price of five commonly traded energy or precious commodities. The energy commodities are the WTI crude oil price index, the US natural gas spot price at the Henry Hub in Louisiana, the average of the Australian coal price at Port Thermal in Newcastle and the South African coal price at Richards Bay. The precious commodities are the gold and silver prices traded in London, afternoon fixing. All the series are sampled in the period February 1994 -- December 2021, and are obtained from the World Bank website \url{https://www.worldbank.org/en/research/commodity-markets}. For each commodity, we  model log returns of their prices, that is  $x_{t,i} = \nabla \log(y_{t,i}) = \log(y_{t,i}/y_{t-1,i})$, where $y_{t,i}$ denotes the price of commodity $i$ ($i=1,\dots,5$) at time $t$. The time plot reported in Figure~\ref{SMfig:1} of the Supplementary Material highlights that the series have  different volatility, which is lower for gold and coal while is more pronounced for oil and natural gas.
\par
We estimate TARMA$(1,1)$ models for the five commodities using   our robust estimation method and the least squares approach,  the latter corresponding to the special case $\alpha = 0$.  We also include the linear ARMA$(1,1)$ model, estimated through full maximum likelihood (ML). In preliminary analyses not reported here, we found estimates for the threshold parameter $r$ consistently close to zero for most values of $\alpha$ ranging from 0 to 1, which  confirms the general asymmetric behavior in growth and contraction periods, see \cite{DL2013}. Motivated by these findings, we set $r=0$ to obtain our final TARMA estimates. Moreover, based on macroeconomic theory, we set $d=1$. The overall model accuracy is assessed through the mean absolute percentage error (MAPE) (see Section~\ref{SMsec:app} if the Supplementary Material), using 12 out-of-sample observations from January to December 2021 as the test set, while the remaining 324 observations are used as the training set.
\par
Table \ref{tab:est resul} shows parameter estimates for the five series with  standard errors  in parentheses below the estimates. The third column shows the values of the tuning parameter $\alpha$, computed by  minimizing the MAPE over a grid of equally spaced values in the interval $(0,1)$. In all the series, we note that the autoregressive and moving-average parameters change, sometimes dramatically, when using our robust method compared to the LS approach. Moreover, the standard errors from the TARMA models based on the LS method are generally larger than the robust standard errors. Thus, using the non-robust method can hinder the discovery of separate regimes and make it impossible to assess the actual significance of many parameters. For instance, for the silver series, the LS method does not show significantly different estimates in the two regimes and its MAPE is even larger than that of the linear ARMA. On the other hand, the robust TARMA reveals the existence of two dynamical regimes with clearly different autoregressive and moving-average behaviors. The  prediction error of the resulting model is 30\% smaller than the LS fit.
\par
\begin{figure}
\centering
\includegraphics[width=0.95\linewidth,keepaspectratio]{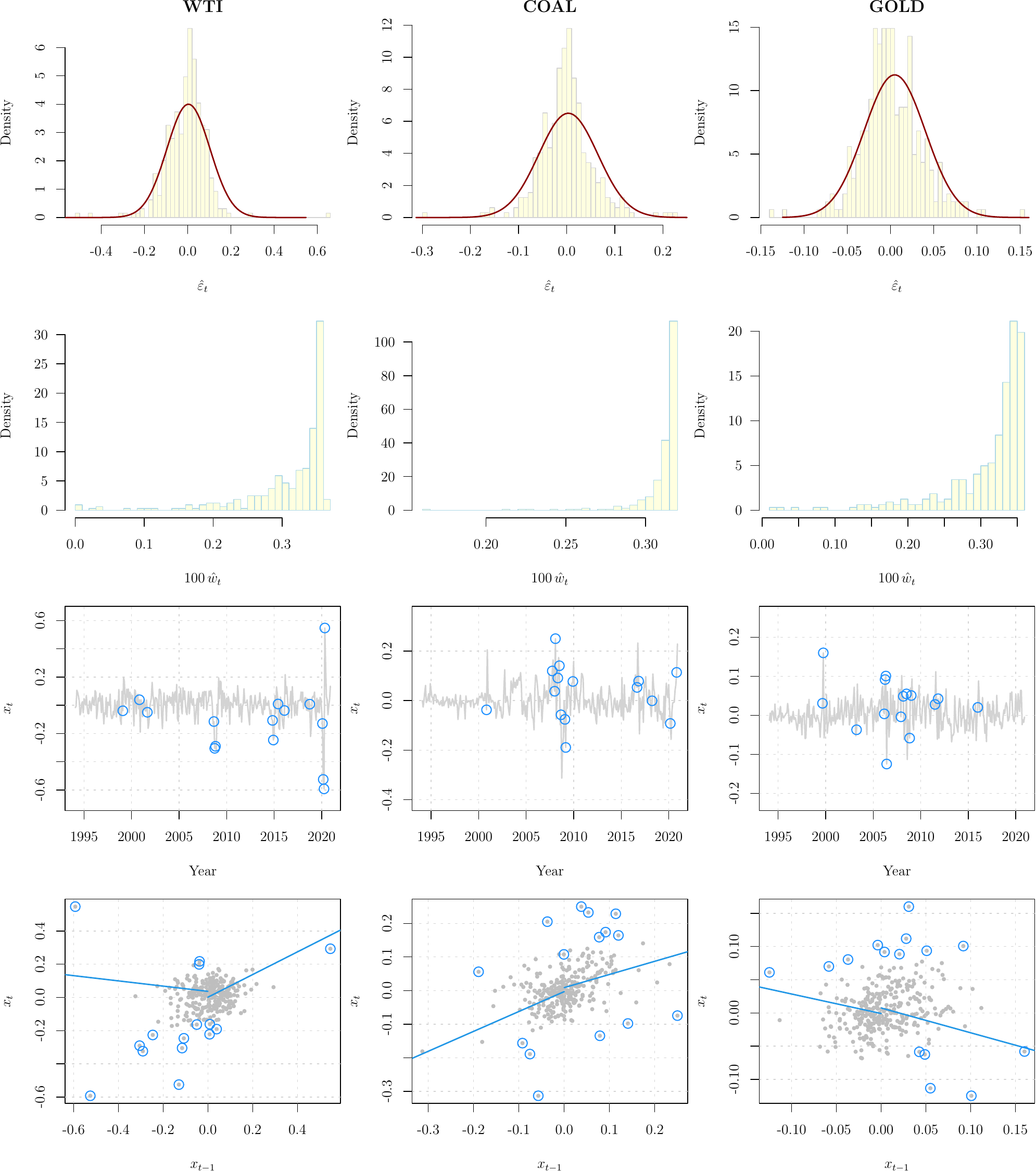}
\caption{Outlier analysis for the crude oil (WTI), coal (COAL) and gold (GOLD) commodity return series. Histograms of the residuals from the robust TARMA fit ($\hat \eps_t$, top row), robust weights ($100 \times \hat w_t$, second row). The 15 most severe outliers are shown by the circles in the time plot of the log-return series ($x_t$, third row), and in the corresponding lag plot of $x_{t}$ versus $x_{t-1}$ (bottom row).} \label{fig:outliers}
\end{figure}

All the estimated intercepts for the robust TARMA are close to zero and this suggests that the transition between the two regimes is not discontinuous (see also Figure~\ref{fig:outliers}, last row). Moreover, in absolute value, the parameters for the lower regime are almost always smaller than those of the upper regime, which highlights the asymmetric behavior of the commodity series characterized by periods of persistent growth and sharp contraction. In particular, the difference in the moving-average parameters denotes the different reaction to shocks in the two regimes: with the exception of coal, in the upper regime the shocks exert a stronger and more persistent influence.
\begin{table}[htbp]
 \caption{TARMA$(1,1)$ parameter estimates for five monthly commodity returns observed in the period March 1997 -- December 2020 based on the least squares (LS) method and the robust method (Rob) with tuning parameter $\alpha$ obtained by minimizing the mean absolute prediction error (MAPE) using the 12 months of 2021 as the test set. The maximum likelihood fit for the linear ARMA$(1,1)$  is added for comparison. Standard errors are reported in parenthesis below their respective estimates.\label{tab:est resul}}
\fontsize{10}{12}\selectfont
\centering
\begin{tabular}[t]{cccccccccc}
\toprule
Series & Model & $\alpha$ & $\phi_{1,0}$ & $\phi_{1,1}$ & $\phi_{2,0}$ & $\phi_{2,1}$ & $\theta_{1,1}$ & $\theta_{2,1}$ & MAPE\\
\midrule
\cellcolor{gray!6}{} & \cellcolor{gray!6}{TARMA (LS)} & \cellcolor{gray!6}{0} & \cellcolor{gray!6}{0.030} & \cellcolor{gray!6}{-0.288} & \cellcolor{gray!6}{-0.006} & \cellcolor{gray!6}{0.552} & \cellcolor{gray!6}{0.848} & \cellcolor{gray!6}{-0.249} & \cellcolor{gray!6}{93.4}\\
 &  &  & (0.013) & (0.229) & (0.008) & (0.212) & (0.187) & (0.197) & \\
\addlinespace
\cellcolor{gray!6}{WTI} & \cellcolor{gray!6}{TARMA (Rob)} & \cellcolor{gray!6}{0.8} & \cellcolor{gray!6}{0.037} & \cellcolor{gray!6}{-0.158} & \cellcolor{gray!6}{0.001} & \cellcolor{gray!6}{0.683} & \cellcolor{gray!6}{0.500} & \cellcolor{gray!6}{-0.589} & \cellcolor{gray!6}{86.9}\\
 &  &  & (0.004) & (0.225) & (0.002) & (0.146) & (0.243) & (0.140) & \\
\addlinespace
\cellcolor{gray!6}{} & \cellcolor{gray!6}{ARMA} & \cellcolor{gray!6}{} & \cellcolor{gray!6}{0.004} & \cellcolor{gray!6}{-0.001} & \cellcolor{gray!6}{} & \cellcolor{gray!6}{} & \cellcolor{gray!6}{0.272} & \cellcolor{gray!6}{} & \cellcolor{gray!6}{97.3}\\
 &  &  & (0.007) & (0.188) &  &  & (0.180) &  & \\
\midrule
\cellcolor{gray!6}{} & \cellcolor{gray!6}{TARMA (LS)} & \cellcolor{gray!6}{0} & \cellcolor{gray!6}{0.001} & \cellcolor{gray!6}{0.647} & \cellcolor{gray!6}{0.043} & \cellcolor{gray!6}{0.111} & \cellcolor{gray!6}{-0.550} & \cellcolor{gray!6}{-0.430} & \cellcolor{gray!6}{100.0}\\
 &  &  & (0.012) & (0.273) & (0.014) & (0.331) & (0.308) & (0.298) & \\
\addlinespace
\cellcolor{gray!6}{NAT GAS} & \cellcolor{gray!6}{TARMA (Rob)} & \cellcolor{gray!6}{0.3} & \cellcolor{gray!6}{0.006} & \cellcolor{gray!6}{0.400} & \cellcolor{gray!6}{0.031} & \cellcolor{gray!6}{-0.500} & \cellcolor{gray!6}{-0.197} & \cellcolor{gray!6}{0.295} & \cellcolor{gray!6}{99.4}\\
 &  &  & (0.002) & (0.035) & (0.002) & (0.039) & (0.033) & (0.033) & \\
\addlinespace
\cellcolor{gray!6}{} & \cellcolor{gray!6}{ARMA} & \cellcolor{gray!6}{} & \cellcolor{gray!6}{0.000} & \cellcolor{gray!6}{-0.136} & \cellcolor{gray!6}{} & \cellcolor{gray!6}{} & \cellcolor{gray!6}{0.178} & \cellcolor{gray!6}{} & \cellcolor{gray!6}{100.0}\\
 &  &  & (0.008) & (0.589) &  &  & (0.583) &  & \\
\midrule
\cellcolor{gray!6}{} & \cellcolor{gray!6}{TARMA (LS)} & \cellcolor{gray!6}{0} & \cellcolor{gray!6}{-0.004} & \cellcolor{gray!6}{0.524} & \cellcolor{gray!6}{0.010} & \cellcolor{gray!6}{0.493} & \cellcolor{gray!6}{-0.214} & \cellcolor{gray!6}{-0.207} & \cellcolor{gray!6}{124.6}\\
 &  &  & (0.005) & (0.172) & (0.006) & (0.172) & (0.176) & (0.189) & \\
\addlinespace
\cellcolor{gray!6}{COAL} & \cellcolor{gray!6}{TARMA (Rob)} & \cellcolor{gray!6}{0.05} & \cellcolor{gray!6}{-0.003} & \cellcolor{gray!6}{0.592} & \cellcolor{gray!6}{0.009} & \cellcolor{gray!6}{0.392} & \cellcolor{gray!6}{-0.317} & \cellcolor{gray!6}{-0.092} & \cellcolor{gray!6}{120.9}\\
 &  &  & (0.000) & (0.010) & (0.000) & (0.015) & (0.009) & (0.016) & \\
\addlinespace
\cellcolor{gray!6}{} & \cellcolor{gray!6}{ARMA} & \cellcolor{gray!6}{} & \cellcolor{gray!6}{0.004} & \cellcolor{gray!6}{0.553} & \cellcolor{gray!6}{} & \cellcolor{gray!6}{} & \cellcolor{gray!6}{-0.169} & \cellcolor{gray!6}{} & \cellcolor{gray!6}{129.2}\\
 &  &  & (0.006) & (0.115) &  &  & (0.135) &  & \\
\midrule
\cellcolor{gray!6}{} & \cellcolor{gray!6}{TARMA (LS)} & \cellcolor{gray!6}{0} & \cellcolor{gray!6}{-0.004} & \cellcolor{gray!6}{0.429} & \cellcolor{gray!6}{0.014} & \cellcolor{gray!6}{0.323} & \cellcolor{gray!6}{-0.472} & \cellcolor{gray!6}{-0.521} & \cellcolor{gray!6}{171.6}\\
 &  &  & (0.003) & (0.297) & (0.004) & (0.276) & (0.291) & (0.267) & \\
\addlinespace
\cellcolor{gray!6}{GOLD} & \cellcolor{gray!6}{TARMA (Rob)} & \cellcolor{gray!6}{0.4} & \cellcolor{gray!6}{-0.001} & \cellcolor{gray!6}{-0.294} & \cellcolor{gray!6}{0.007} & \cellcolor{gray!6}{-0.373} & \cellcolor{gray!6}{0.371} & \cellcolor{gray!6}{0.505} & \cellcolor{gray!6}{104.7}\\
 &  &  & (0.000) & (0.063) & (0.000) & (0.045) & (0.061) & (0.040) & \\
\addlinespace
\cellcolor{gray!6}{} & \cellcolor{gray!6}{ARMA} & \cellcolor{gray!6}{} & \cellcolor{gray!6}{0.005} & \cellcolor{gray!6}{-0.247} & \cellcolor{gray!6}{} & \cellcolor{gray!6}{} & \cellcolor{gray!6}{0.398} & \cellcolor{gray!6}{} & \cellcolor{gray!6}{135.6}\\
 &  &  & (0.002) & (0.252) &  &  & (0.236) &  & \\
\midrule
\cellcolor{gray!6}{} & \cellcolor{gray!6}{TARMA (LS)} & \cellcolor{gray!6}{0} & \cellcolor{gray!6}{0.007} & \cellcolor{gray!6}{-0.068} & \cellcolor{gray!6}{0.000} & \cellcolor{gray!6}{-0.039} & \cellcolor{gray!6}{0.302} & \cellcolor{gray!6}{0.338} & \cellcolor{gray!6}{122.7}\\
 &  &  & (0.008) & (0.338) & (0.008) & (0.329) & (0.310) & (0.321) & \\
\addlinespace
\cellcolor{gray!6}{SILVER} & \cellcolor{gray!6}{TARMA (Rob)} & \cellcolor{gray!6}{0.8} & \cellcolor{gray!6}{0.004} & \cellcolor{gray!6}{0.348} & \cellcolor{gray!6}{0.003} & \cellcolor{gray!6}{-0.681} & \cellcolor{gray!6}{-0.178} & \cellcolor{gray!6}{0.875} & \cellcolor{gray!6}{89.6}\\
 &  &  & (0.000) & (0.028) & (0.001) & (0.025) & (0.026) & (0.019) & \\
\addlinespace
\cellcolor{gray!6}{} & \cellcolor{gray!6}{ARMA} & \cellcolor{gray!6}{} & \cellcolor{gray!6}{0.005} & \cellcolor{gray!6}{-0.072} & \cellcolor{gray!6}{} & \cellcolor{gray!6}{} & \cellcolor{gray!6}{0.299} & \cellcolor{gray!6}{} & \cellcolor{gray!6}{106.0}\\
 &  &  & (0.004) & (0.176) &  &  & (0.164) &  & \\
\bottomrule
\end{tabular}

\end{table}%
As already mentioned, the higher estimation accuracy of the robust TARMA is also beneficial for prediction since this model always outperforms the least square TARMA. Gains in terms of MAPE are sizeable for precious commodities (up to 67\% for gold and 33\% for silver); they are moderate for coal (4\%) and oil (6\%), and small for natural gas (1\%). The linear ARMA model is the least accurate, except for gold, where many parameter estimates are not significant.
\par
In order to detect the most influential outliers, for each series we  compute the robust weights
$$
\hat w_t = \dfrac{\exp\{ - \hat \alpha \times \hat \eps^2_{t}/ (2\hat \sigma^2)\}}{\sum_{s=1}^n\exp\{ - \hat \alpha \times \hat \eps^2_{s}/ (2\hat \sigma^2)\}}, \ \ t=1,\dots, 324,
$$
where $\hat \eps_t$ is the  residual at time $t$ from the robust fit, $\hat \sigma^2$ is the estimated error variance and $\hat \alpha$ is the data-driven tuning parameter obtained by minimizing the MAPE. Smaller weights correspond to observations that are further from the assumed clean model, i.e. the TARMA model with Gaussian errors.

Figure~\ref{fig:outliers} (top) shows the histograms of the residuals from the robust TARMA for oil, coal and gold prices. All the histograms appear to be different from the nominal standard Gaussian density (superimposed in red) due to heavy tails or asymmetry. The residuals translate into robust weights mostly concentrated on larger values above $0.30$, although a number of observations receives smaller weights closer to zero (see Figure~\ref{fig:outliers}, second row), indicating the presence of strong outliers. Figure~\ref{fig:outliers} (third row) highlights with circles the most influential outliers corresponding to the smallest robust weights (5\% of the sample, or 15 values) in the time plot of the original log-return series. Many of these extreme outliers are evident and correspond to the shocks which occurred during the financial crisis in 2008--2009 and at the beginning of the COVID pandemic, although some of the abrupt changes appear to be compatible with the assumed TARMA model.
\par
The last row of Figure~\ref{fig:outliers} shows the same outliers in the state space (lag plot of $x_{t}$ versus  $x_{t-1}$), where we have also added  estimated piecewise linear autoregression lines. Note that the 15 most extreme observations are, in fact, outliers in the state space while this is not so evident from the time plot (Figure~\ref{fig:outliers}, third row). Moreover, the placement of such observations marked as outliers appears to be linked to the  commodity type. For precious commodities (gold and silver) the outliers tend to fall in the upper regime, while for oil they are found in the lower regime. Finally, for gas and coal, there is roughly the same proportion of outliers in both regimes. Figure~\ref{SMfig:2} in the Supplementary Material reports the plots for the two remaining commodities (oil and silver).

\section{Conclusions}\label{sec:concl}

TARMA models have attracted considerable interest due to their ability to parsimoniously describe complex dynamical features such as jumps, asymmetric limit cycles, time irreversibility, and chaos. They are unique in that they provide a natural interpretation for phenomena that change qualitatively across regimes and react differently to shocks. Nonetheless, estimation for TARMA model is currently limited to the least squares method, which is known to be severely influenced by the presence of outliers. In this paper we provide the first theoretical framework for robust M-estimation for TARMA models and also study its practical relevance.
\par
Theorems \ref{thm:cons} and \ref{thm:rate} extend the results of \cite{Li11a} and establish an asymptotic theory for a wide class of estimators found as the solution of M-estimating equations with bounded derivatives. We establish the superconsistency for $\hat r_n$ and defer to future research the derivation of the limit distribution of the threshold estimator, which is a challenging task. Our results can be used to derive other robust inference and model selection tools for TARMA processes. For example, following \cite{ronchetti1997robustness} and \cite{muller2009robust}, a robust model-selection criterion for TARMA models may be formulated as $2 {\rho}_n(\hat \bEta_n) + 2 \text{trace}(\hat \bH^{-1} \hat \bJ)$, where $\hat \bH$ and $\hat \bJ$ are the plug-in estimators based on the sensitivity and variability estimators defined in (\ref{eqn:sens_var}). These can also be used to derive Wald and score statistics to test hypotheses on the parameters. Focusing on the re-descending estimator of \cite{Fer12}, we study the robustness properties of the proposed M-estimator in a range of scenarios involving both additive and innovation outliers. The results from our Monte Carlo experiments show that moving away from the LS estimator even by a small amount already achieves robustness both in terms of bias and variance. Overall, our estimator reduces considerably the asymptotic bias also in the presence of severe contaminations and high fractions of outliers, where the least squares estimator fails. The findings suggest that robust M-estimation should be generally preferred to the least squares method, even when the actual data deviate only slightly from the nominal TARMA model.
\par
The analysis of the five time series of commodity prices shows that the robust TARMA estimates present smaller standard errors and lead to superior forecasting accuracy compared to the least squares fit. This enables us to detect regime changes with confidence and support the hypothesis of a two-regime, asymmetric nonlinearity around zero, characterised by slow expansions and fast contractions. Although a thorough analysis of the price dynamics for different commodities is beyond the scope of the present work, the robust TARMA framework could be used as the foundation for future modelling approaches, possibly leading to important advancements in the field.
\par
An interesting direction for future investigations could be the study of the performance in the presence of specific contamination processes. For example IOs are generally more challenging to handle and would require the development of some ad-hoc estimating function. One possible approach is to introduce a robust filtering step within the residual function, as in the bounded innovation propagation ARMA (BIP-ARMA) of \cite{muler2009robust}. For the time being, we note that there is a fundamental difference in the way non-linear processes react to perturbations compared to linear processes. In general, the presence of dynamic noise can alter qualitatively and non trivially the nature of the process, see e.g.,  \cite{Cha01} for a discussion. For instance, for linear processes the response function to noise is flat, whereas non-linear processes can act both as noise amplifiers and noise suppressors, producing a plethora of characteristic phenomena, such as resonances, or the state-dependence predictability, which is well known in the forecasting literature, see e.g., \cite{Fan05}, Ch. 10.


\newpage
\setcounter{table}{0}
\setcounter{section}{0}
\setcounter{page}{1}
\renewcommand{\thesection}{\Alph{section}}

\begin{center}
\section*{\Large Supplement for:\\ Robust estimation for Threshold  Autoregressive Moving-Average models}
\bookmark[level=0, page=18]{Supplementary Material}
\end{center}

\bigskip

\begin{center}
\Large
{Greta Goracci, Davide Ferrari, Simone Giannerini, Francesco Ravazzolo}
\end{center}
\bigskip
\bigskip

In Section~\ref{SMsec:proofs} we report  the technical proofs leading to Theorems 1 and 2 of the main paper. Section~\ref{SMsec:app} contains supplementary figures related to the analysis of the set of commodity time series.

\section{Proofs}\label{SMsec:proofs}

For clarity of presentation and without loss of generality the proofs are detailed for the case $p=q=1$. Analogous arguments hold for the general case $p,q>1$. In order to prove Theorems \ref{thm:cons} and \ref{thm:rate} we need some technical lemmas. For each $\bEta\in\mathcal{H}$ and $0<\delta<1$ define the neighborhood
\begin{align}
  U_\delta(\bEta)=\left\{\bEta^*=(\bphi^*,\bth^*,r^*,d^*)\in \mathcal{H}:\right.& \left\|\bphi^*-\bphi\right\|\leq \delta, \left\|\bth^*-\bth\right\|\leq \delta,\nonumber\\
  &\left. |r^*-r|\leq \delta\right\}\label{eq:Udelta}.
\end{align}
\begin{lemma}\label{lem:continuity}
  For each $\bEta\in\mathcal{H}$ it holds that
  $$\sup_{\bEta^*\in U_\delta(\bEta)}E\left[\rho(\eps_t(\bEta^*))-\rho(\eps_t(\bEta))\right]\xrightarrow[]{}0,\quad\text{as}\quad \delta\to0,$$
\end{lemma}
with $\eps_t(\cdot)$ being defined in Equation~(\ref{eq:res}).

\subsubsection*{Proof of Lemma \ref{lem:continuity} }

  We exploit the following representation for TARMA models due to \citet{Lin05}, Theorem A.1 and A.2. For any $\bEta\in\mathcal{H}$,  the residual function can be represented as follows:
  \begin{align*}
  \eps_t(\bEta)=\sum_{j=0}^{\infty}H_{t,j}(\bEta)\times \left[X_{t-j}-\right.&\left\{\phi_{1,0}+ \phi_{1,1}X_{t-1-j}\right\}I(X_{t-d-j}\leq r)\\
  -&\left.\left\{\phi_{2,0}+ \phi_{2,1}X_{t-1-j}\right\}I(X_{t-d-j}> r)\right],
  \end{align*}
  where $$H_{t,j}(\bEta)=\left\{\theta_{2,1}+(\theta_{1,1}-\theta_{2,1})I(X_{t-d}\leq r)\right\}^j.$$
  Moreover, there exists a positive $\gamma<1$ such that
  \begin{equation}\label{eq:TARMA_repres}
\sup_{\bEta\in\mathcal{H}}\left\|H_{t,j}(\bEta)\right\|=O(\gamma^j),
  \end{equation}
  Using some straightforward algebra, we compute the difference
  \begin{align}
  &\eps_t(\bEta^*)-\eps_t(\bEta)\nonumber\\
  &=\sum_{j=0}^{\infty}H_{t,j}(\bEta^*)\left[X_{t-j}-\left\{\phi^*_{1,0}+ \phi^*_{1,1}X_{t-1-j}\right\}I(X_{t-d-j}\leq r^*)-\left\{\phi^*_{2,0}+ \phi^*_{2,1}X_{t-1-j}\right\}I(X_{t-d-j}> r^*)\right]\nonumber\\
  &-\sum_{j=0}^{\infty}H_{t,j}(\bEta^*)\left[X_{t-j}-\left\{\phi_{1,0}+ \phi_{1,1}X_{t-1-j}\right\}I(X_{t-d-j}\leq r)-\left\{\phi_{2,0}+ \phi_{2,1}X_{t-1-j}\right\}I(X_{t-d-j}> r)\right]\nonumber\\
  &+\sum_{j=0}^{\infty}H_{t,j}(\bEta^*)\left[X_{t-j}-\left\{\phi_{1,0}+ \phi_{1,1}X_{t-1-j}\right\}I(X_{t-d-j}\leq r)-\left\{\phi_{2,0}+ \phi_{2,1}X_{t-1-j}\right\}I(X_{t-d-j}> r)\right]\nonumber\\
  &-\sum_{j=0}^{\infty}H_{t,j}(\bEta)\left[X_{t-j}-\left\{\phi_{1,0}+ \phi_{1,1}X_{t-1-j}\right\}I(X_{t-d-j}\leq r)-\left\{\phi_{2,0}+ \phi_{2,1}X_{t-1-j}\right\}I(X_{t-d-j}> r)\right]\nonumber\\
  &=\sum_{j=0}^{\infty}H_{t,j}(\bEta^*)\left[\bphi^\intercal\cdot\nabla_r(X_{t-1-j})-\bphi^{*\intercal}\cdot\nabla_{r^*}(X_{t-1-j})\right] \label{lem1.a}\\
  &+\sum_{j=0}^{\infty}\left[H_{t,j}(\bEta^*)-H_{t,j}(\bEta)\right]\left[X_{t-j}- \bphi^\intercal\cdot\nabla_r(X_{t-1-j})\right]\label{lem1.b},
  \end{align}
  where
  $$\nabla_r(X_{t})=\left(I(X_{t-d+1}\leq r),X_{t}I(X_{t-d+1}\leq r),I(X_{t-d+1}>r),X_{t}I(X_{t-d+1}>r)\right)^\intercal.$$
  Note that
  \begin{align*}
  \left\|\nabla_r(X_{t})\right\|&=\sqrt{1+X_{t}^2}\leq 1+|X_{t}|,\\
  \left\|\nabla_r(X_{t})-\nabla_{r^*}(X_{t})\right\|&=\sqrt{2(1+X_{t}^2)}I(r\wedge r^*<X_{t-d+1}\leq r\vee r^*)\\
  &\leq \sqrt{2}(1+|X_{t}|) I(|X_{t-d+1}-r|\leq |r-r^*|),\\
  \left|\bphi^\intercal\cdot\nabla_r(X_{t})- \bphi^{*\intercal}\cdot\nabla_{r^*}(X_{t})\right|&=\left|\bphi^\intercal\cdot\left(\nabla_r(X_{t})-\nabla_{r^*}(X_{t})\right) + \left(\bphi^\intercal-\bphi^{*\intercal}\right)\nabla_{r^*}(X_{t})\right|\\
  &\leq \sqrt{2}(1+|X_{t}|)\left[\left\|\bphi \right\| I(|X_{t-d+1}-r|\leq |r-r^*|)+\left\|\bphi-\bphi^*\right\|\right].
  \end{align*}
  By the Cauchy-Schwarz inequality for the first term in (\ref{lem1.a}) we have
  \begin{align*}
  &\left|\sum_{j=0}^{\infty}H_{t,j}(\bEta^*)\left[\bphi^\intercal\cdot\nabla_r(X_{t-1-j})-\bphi^{*\intercal}\cdot\nabla_{r^*}(X_{t-1-j})\right]\right|\\
  &\leq\sum_{j=0}^{\infty}\left\|H_{t,j}(\bEta^*)\right\|\left|\bphi^\intercal\cdot\nabla_r(X_{t-1-j})-\bphi^{*\intercal}\cdot\nabla_{r^*}(X_{t-1-j})\right|\\
  &\leq K \sum_{j=0}^{\infty}\gamma^j\left\{\sqrt{2}(1+|X_{t-1-j}|)\left[\left\|\bphi \right\| I(|X_{t-d-j}-r|\leq \delta)+\delta\right]\right\}:=\Upsilon^{(1)}_\delta(\mathcal{F}_{t-1}),
  \end{align*}
  with $K$ being a positive constant. Here $\Upsilon^{(1)}_\delta(\mathcal{F}_{t-1})$ indicates that the function $\Upsilon^{(1)}_\delta$ depends upon $X_{t-1},X_{t-2},\dots$.

By Assumption (A1) and the argument in  Lemma 3.1 in \citet{Kou03}, it holds that $E[\Upsilon^{(1)}_\delta(\mathcal{F}_{t-1})] \to 0$ as $\delta\to0$. An analogous calculation shows that the absolute value of (\ref{lem1.b}) is bounded by a function, say $\Upsilon^{(2)}_\delta(\mathcal{F}_{t-1})$, such that $E[\Upsilon^{(2)}_\delta(\mathcal{F}_{t-1})]\to 0$  as $\delta\to0$. Letting $\Upsilon_\delta(\mathcal{F}_{t-1})=\Upsilon^{(1)}_\delta(\mathcal{F}_{t-1})+\Upsilon^{(2)}_\delta(\mathcal{F}_{t-1})$, we get
$|\eps_t(\bEta^*)- \eps_t(\bEta)|\leq \Upsilon_\delta(\mathcal{F}_{t-1})$
with $E\left[\Upsilon_\delta(\mathcal{F}_{t-1})\right]\to 0$ as $\delta \to 0$.  Finally, Assumption (A3) implies that there exists a constant $\mathcal{K}$ such that
  \begin{align*}
  E\left[\sup_{\bEta^* \in U_{\delta}(\bEta)}|\rho(\eps_t(\bEta^*))-\rho(\eps_t(\bEta))|\right] \leq E\left[\int_{-\Upsilon_\delta(\mathcal{F}_{t-1})}^{\Upsilon_\delta(\mathcal{F}_{t-1})} |\rho'(\eps_t(\bEta)+w)|dw\right]
  \leq \mathcal{K} \times E\left[\Upsilon_\delta(\mathcal{F}_{t-1})\right].
  \end{align*}
\noindent
The right hand side of the above expression goes to zero as $\delta\to 0$, which completes the proof.
 \subsubsection*{Proof of Theorem \ref{thm:cons}}
  We prove that for any neighborhood $\mathcal{U}$ of $\bEta_0$, for any sufficiently large $n$,
    $$\inf_{\bEta\in\mathcal{U}^c}\left\{\brho_n(\bEta)-\brho_n(\bEta_0)\right\}> \inf_{\bEta\in\mathcal{U}}\left\{\brho_n(\bEta)-\brho_n(\bEta_0)\right\}, \quad\text{a.s.}$$
  Since
  $$
  \inf_{\bEta\in\mathcal{U}}\frac{1}{n}\left\{\brho_n(\bEta)-\brho_n(\bEta_0)\right\}\leq\frac{1}{n}\left\{\brho_n(\bEta_0)-\brho_n(\bEta_0)\right\}=0,
  $$
  it suffices to show
  \begin{equation}\label{eq:cons_main}
    \inf_{\bEta\in\mathcal{U}^c}\frac{1}{n}\left\{\brho_n(\bEta)-\brho_n(\bEta_0)\right\}> 0.
  \end{equation}
Consider the expectations
  \begin{align*}
  g(\bEta)=E\left[\rho(\eps_t(\bEta))-\rho(\eps_t(\bEta_0))\right]&=E\left[\rho(\eps_t(\bEta_0)+\left\{\eps_t(\bEta)-\eps_t(\bEta_0)\right\})- \rho(\eps_t(\bEta_0))\right],\\
  \text{and}\quad l(x)&= E\left[\rho(\eps_t(\bEta_0)+x)-\rho(\eps_t(\bEta_0))\right].
  \end{align*}
  \noindent
  Note that the law of iterated expectations implies that $g(\bEta)=E[l(\eps_t(\bEta)-\eps_t(\bEta_0))]$. Moreover, following the same argument as in Lemma 6.4 of \citet{Li13}, it is not difficult to show that, for any $\bEta\neq\bEta_0$, there exists $x_0>0$ such that
  \begin{equation}\label{eqn:pos}
  P(|\eps_t(\bEta)-\eps_t(\bEta_0)|>x_0)>0.
  \end{equation}
  By Assumption (A3), $l(x)>0$ for any $x\neq0$, thereby (\ref{eqn:pos}) implies that $g(\bEta)=0$ if $\bEta=\bEta_0$ and it is strictly positive if $\bEta\neq\bEta_0$. Hence,  for any neighborhood $\mathcal{U}$  of $\bEta_0$ there exists $\tilde{\bEta}\in\mathcal{U}^c$ such that
  \begin{equation*}
    \inf_{\bEta\in\mathcal{U}^c}g(\bEta)=g(\tilde{\bEta})>0.
  \end{equation*}
  Note that Lemma~\ref{lem:continuity} implies that for all $c>0$
  \begin{equation}\label{eq:def_lim}
     \exists\; \bar{\delta}:\quad \forall \delta<\bar{\delta}\qquad \left|\sup_{\bEta^*\in U_{\delta}(\bEta)} E\left[\rho(\eps_t(\bEta^*))-\rho(\eps_t(\bEta))\right]\right|<c.
  \end{equation}
  We consider the neighborhood $U_{\bar\delta}$ and prove that:
  \begin{equation}\label{eq:bound1}
    E\left[\inf_{\bEta^*\in U_{\bar\delta}(\bEta)}\left\{\rho(\eps_t(\bEta^*))-\rho(\eps_t(\bEta_0))\right\}\right]\geq 2c \qquad \forall \bEta\in\mathcal{U}^c.
  \end{equation}
  To this end consider $\bEta\in\mathcal{U}^c$; it holds that
  \begin{align*}
  &E\left[\inf_{\bEta^*\in U_{\bar\delta}(\bEta)}\left\{\rho(\eps_t(\bEta^*))-\rho(\eps_t(\bEta_0))\right\}\right]\\
  &=E\left[\inf_{\bEta^*\in U_{\bar\delta}(\bEta)}\left\{\rho(\eps_t(\bEta^*))-\rho(\eps_t(\bEta_0)) +\rho(\eps_t(\bEta))-\rho(\eps_t(\bEta))\right\}\right]\\
  &\geq E\left[\rho(\eps_t(\bEta))-\rho(\eps_t(\bEta_0))\right]- E\left[ \sup_{\bEta^*\in U_{\bar\delta}(\bEta)}\left\{ \rho(\eps_t(\bEta^*))-\rho(\eps_t(\bEta))\right\}\right]\\
  &\geq \inf_{\bEta\in\mathcal{U}^c} E\left[\rho(\eps_t(\bEta))-\rho(\eps_t(\bEta_0))\right]-c=2c,
  \end{align*}
  where the last equality holds by setting $c=g(\tilde{\bEta})/3$ in (\ref{eq:def_lim}). Since $\mathcal{U}^c$ is compact, there exists a finite coverage $\{U_{\bar{\delta}}(\bEta_\kappa),\;\kappa=1,\dots,K\}$, with $K$ being a constant, such that $\bEta_\kappa\in\mathcal{U}^c$, for each $\kappa=1,\dots,K$, and $\bigcup_{\kappa=1}^{K}U_{\bar{\delta}}(\bEta_\kappa)=\mathcal{U}^c$. The ergodicity of $\{X_t\}$ implies that, for any $\bEta_\kappa$:
  \begin{align*}
  &\inf_{\bEta\in U_{\bar\delta}(\bEta_\kappa)}\frac{1}{n}\sum_{t=1}^{n}\left\{\rho(\eps_t(\bEta_0)+\left\{\eps_t(\bEta)-\eps_t(\bEta_0)\right\})-\rho(\eps_t(\bEta_0))\right\}\\ \xrightarrow[n\to\infty]{a.s.}&\quad E\left[\inf_{\bEta\in U_{\bar\delta}(\bEta_\kappa)}\left\{\rho(\eps_t(\bEta_0)+\left\{\eps_t(\bEta)-\eps_t(\bEta_0)\right\})-\rho(\eps_t(\bEta_0))\right\}\right],
  \end{align*}
  thereby for all $c>0$ there exists $\bar{n}$ such that for any $n>\bar{n}$
  \begin{equation}\label{eq:def_lim2}
    \left|\inf_{\bEta\in U_{\bar\delta}(\bEta_\kappa)}\frac{1}{n}\left\{\brho_n(\bEta)-\brho_n(\bEta_0)\right\}- E\left[\inf_{\bEta\in U_{\bar\delta}(\bEta_\kappa)}\left\{\rho(\eps_t(\bEta_0)+\left\{\eps_t(\bEta)-\eps_t(\bEta_0)\right\})-\rho(\eps_t(\bEta_0))\right\}\right]\right|<c.
  \end{equation}
  In particular, (\ref{eq:bound1}) implies
  \begin{align*}
  \inf_{\bEta\in U_{\bar\delta}(\bEta_\kappa)}\frac{1}{n}\left\{\brho_n(\bEta)-\brho_n(\bEta_0)\right\}&\geq E\left[\inf_{\bEta\in U_{\bar\delta}(\bEta_\kappa)}\left\{\rho(\eps_t(\bEta_0)+\left\{\eps_t(\bEta)-\eps_t(\bEta_0)\right\})-\rho(\eps_t(\bEta_0))\right\}\right] -c \\
  &\geq2c-c=c.
  \end{align*}
Hence, (\ref{eq:cons_main}) is satisfied and the proof is complete.
\subsubsection*{Proof of Theorem \ref{thm:rate}}
Since $\hat\bEta_n$ is consistent for the population parameter $\bEta_0$ defined in Section \ref{sec:properties}, we restrict the parameter space $\mathcal{H}$ to the neighborhood
$\mathcal{H}_\delta=\left\{\bEta\in\mathcal{H}:\left\|\blam-\blam_0\right\|<\delta\wedge |r-r_0|<\delta\right\}$,
with $0<\delta<1$ to be determined later. Without loss of generality, we can assume $r=r_0+u$ with $u$ being a positive real value.
\par\noindent
\textbf{(i)} To show the first part,  we proceed similarly to \cite{Li11a}. The result follows by showing that $\forall c>0$, $\exists\; \beta,B>0$ such that $\forall n$ sufficiently large:
\begin{equation}\label{eq:rate_r_main}
  P\left(\inf_{\begin{array}{c}
                 \bEta\in \mathcal{H}_\delta \\
                 B/n<r-r_0<\delta
               \end{array}}\frac{\brho_n(\blam,r)-\brho_n(\blam,r_0)}{nP(r_0<X_{t-1}\leq r)}>\beta\right)>1-c.
\end{equation}
\noindent
See \citet{Kou03}, proof of Theorem 3.2.\par
Define
\begin{align*}
D_n(\blam,r)&=\brho_n(\blam,r)-\brho_n(\blam,r_0)=D_n^{(1)}(r)+ D_n^{(2)}(\blam,r),
\end{align*}
where
\begin{align*}
D_n^{(1)}(r)&= \brho_n(\blam_0,r)-\brho_n(\blam_0,r_0),\\
D_n^{(2)}(\blam,r)&=[\brho_n(\blam,r)-\brho_n(\blam,r_0)]-[\brho_n(\blam_0,r_0)-\brho_n(\blam_0,r_0)].
\end{align*}
We start focusing on $D_n^{(1)}(r)$. Let $\mathbf{Z}_t=(X_{t},\eps_t)^\intercal$ and $\mathbf{Z}^*=(\mathbf{a}^\intercal,\mathbf{b}^\intercal)^\intercal$, with $\mathbf{a}$ and $\mathbf{b}$ being defined in Assumption (A4). Moreover we set $d_i$, $i=1,2,3$ to be the differences between the corresponding true parameters in the two regimes, i.e. $d_0=(\phi_{0,1,0}-\phi_{0,2,0})$, $d_1=(\phi_{0,1,1}-\phi_{0,2,1})$  and $d_2=(\theta_{0,1,1}-\theta_{0,2,1})$.

   Then  there exists a constant ${C}>0$ such that
 $$d_{t-1}:= (d_0,d_1,d_2)\cdot(1,\mathbf{Z}^{\intercal})^\intercal= d_0+d_1X_{t-1}+d_2\eps_{t-1}$$
 is bounded away from zero for any $\mathbf{Z}_t$ satisfying $\|\mathbf{Z}_{t-1}-\mathbf{Z}^*\|\leq {C}$. Moreover, let
\begin{align*}
I^*(r_0<X_{t}\leq r)&=I(r_0<X_{t}\leq r \wedge \|\mathbf{Z}_t-\mathbf{Z}^*\|\leq C),\\
P^*(r_0<X_{t}\leq r)&=P(r_0<X_{t}\leq r \wedge \|\mathbf{Z}_t-\mathbf{Z}^*\|\leq C).
\end{align*}
\noindent
Routine algebra implies that:
\begin{align*}
\frac{D_n^{(1)}(r)}{nP(r_0<X_{t-1}\leq r)}&=\frac{1}{nP(r_0<X_{t-1}\leq r)}\sum_{t=1}^{n}\left[\rho(\eps_t+\left\{\eps_t(\blam_0,r)-\eps_t\right\})-\rho(\eps_t)\right]\\
&\geq \Gamma^{(1)}_n+\Gamma^{(2)}_n+\Gamma^{(3)}_n,
\end{align*}
with
\begin{align*}
\Gamma^{(1)}_n&=\frac{1}{nP(r_0<X_{t-1}\leq r)}\sum_{t=1}^{n}\left[\rho(\eps_t+d_{t-1})-\rho(\eps_t)\right]I^*(r_0<X_{t-1}\leq r),\\
\Gamma^{(2)}_n&=\frac{1}{nP(r_0<X_{t-1}\leq r)}\sum_{t=1}^{n}\left[\rho(\eps_t+\left\{\eps_t(\blam_0,r)-\eps_t\right\})- \rho(\eps_t+d_{t-1})\right]I^*(r_0<X_{t-1}\leq r),\\
\Gamma^{(3)}_n&=\frac{1}{nP(r_0<X_{t-1}\leq r)}\sum_{t=1}^{n}\left[\rho(\eps_t+\left\{\eps_t(\blam_0,r)-\eps_t\right\})- \rho(\eps_t)\right]\left\{P^*(r_0<X_{t-1}\leq r) - I^*(r_0<X_{t-1}\leq r)\right\}.
\end{align*}
First we show that $\Gamma^{(2)}_n$ and $\Gamma^{(3)}_n$  are negligible in probability. To this end note that, when $r_0<X_{t-1}\leq r$ then
$$\eps_t(\blam_0,r)-\eps_t-d_{t-1}=-\left\{\theta_{0,2,1}+(\theta_{0,1,1}-\theta_{0,2,1})I(X_{t-1}\leq r)\right\}\left\{\eps_{t-1}(\blam_0,r)-\eps_{t-1}\right\}.$$
By Assumption A3, we have
\begin{align*}
&\left|\rho(\eps_t+\left\{\eps_t(\blam_0,r)-\eps_t\right\})- \rho(\eps_t+d_{t-1})\right|\\
&\leq\int_{-|\eps_t(\blam_0,r)-\eps_t-d_{t-1}|}^{|\eps_t(\blam_0,r)-\eps_t-d_{t-1}|}|\rho'(\eps_t+y)|dy\\
&\leq K_1|\eps_t(\blam_0,r)-\eps_t-d_{t-1}|\\
&\leq K_1 I(r_0<X_{t-1}\leq r)\sum_{j=0}^{\infty}\gamma^j\|\mathbf Z_{t-2-j}\|I(r_0<X_{t-2-j}\leq r),
\end{align*}
with $K_1$ being a positive constant. The last inequality follows from the TARMA representation and (\ref{eq:TARMA_repres}). Hence $\Gamma^{(2)}_n$ is negligible in probability by Part A.2 of Lemma A.1 in  \citet{Li11a}. By a similar argument, we have
\begin{align*}
|\Gamma^{(3)}_n|&\leq \frac{K_2}{P(r_0<X_{t-1}\leq r)}\frac{1}{n}\sum_{t=1}^{n}\left|I^*(r_0<X_{t-1}\leq r)-P^*(r_0<X_{t-1}\leq r)\right|
\end{align*}
for some $K_2>0$,  which is negligible in probability by  Part A.1 of Lemma A.1 in \citet{Li11a}. Now, we focus on $\Gamma^{(1)}_n$. Since $\|\mathbf{Z}_{t-1}-\mathbf{Z}^*\|\leq C$, by Assumption (A3), we have $\rho(\eps_t+d_{t-1})-\rho(\eps_t):=c_0>0$ and
 \begin{align*}
 \Gamma^{(1)}_n&>c_0\frac{P^*(r_0<X_{t-1}\leq r)}{{P}(r_0<X_{t-1}\leq r)}\frac{1}{n}\sum_{t=1}^{n}\frac{I^*(r_0<X_{t-1}\leq r)}{P^*(r_0<X_{t-1}\leq r)}
 \end{align*}
 By  Lemma A.1 in \citet{Li11a} and
 $$\lim_{u\to0}\frac{P^*(r_0<X_{t-1}\leq r_0+u)}{{P}(r_0<X_{t-1}\leq r_0+u)}>0,$$
 \noindent
 there exists a sufficiently small $\delta>0$ such that for all $c>0$ there exist $\beta_1,B>0$ such that for any sufficiently large  $n$, Equation~(\ref{eq:rate_r_main}) is satisfied.

As concerns $D^{(2)}_{n}(\blam,r)$ note that
$$\frac{D^{(2)}_{n}(\blam,r)}{n}= \frac{1}{n}\sum_{t=1}^{n}\int_{0}^{1}\left\{\frac{\partial\rho(\blam_0+w(\blam-\blam_0),r)}{\partial\blam} - \frac{\partial\rho(\blam_0+w (\blam-\blam_0),r_0)}{\partial\blam}\right\}(\blam-\blam_0)dw$$
and it can be shown that
$$\sup_{\begin{array}{c}
                 \bEta\in \mathcal{H}_\delta \\
                 B/n<r-r_0<\delta
               \end{array}}\frac{|D^{(2)}_{n}(\blam,r)|}{nP(r_0<X_{t-1}\leq r)}=O_p(\delta).$$
The proof is completed by noting that
\begin{align*}
&\inf_{\begin{array}{c}
                 \bEta\in \mathcal{H}_\delta \\
                 B/n<r-r_0<\delta
               \end{array}}\frac{\brho_n(\blam,r)-\brho_n(\blam,r_0)}{nP(r_0<X_{t-1}\leq r)}\\
&\geq \inf_{\begin{array}{c}
                 B/n<r-r_0<\delta
               \end{array}}\frac{D^{(1)}_n(r)}{nP(r_0<X_{t-1}\leq r)} - \sup_{\begin{array}{c}
                 \bEta\in \mathcal{H}_\delta \\
                 B/n<r-r_0<\delta
               \end{array}}\frac{|D^{(2)}_{n}(\blam,r)|}{nP(r_0<X_{t-1}\leq r)}.
\end{align*}

\par\noindent
\textbf{(ii)} Consider the first order Taylor's expansion of $\partial\brho_n(\blam,r)/\partial\blam$:
\begin{equation}\label{eq:Taylor}
  \frac{\partial\brho_n(\hat{\blam}_n(r),r)}{\partial\blam}= \frac{\partial\brho_n({\blam}_0,r)}{\partial\blam} + \frac{\partial^2\brho_n(\bar{\blam},r)}{\partial\blam\partial\blam^\intercal}(\hat{\blam}_n(r)-\blam_0),
\end{equation}
with $\bar{\blam}$ being between $\hat{\blam}_n(r)$ and ${\blam}_0$. Hence
$$\frac{1}{n}\frac{\partial\brho_n({\blam}_0,r)}{\partial\blam} + \frac{1}{n}\frac{\partial^2\brho_n(\bar{\blam},r)}{\partial\blam\partial\blam^\intercal}(\hat{\blam}_n(r)-\blam_0)=0.$$
The ergodicity of $\{X_t\}$ implies that
$$\frac{1}{n}\frac{\partial^2\brho_n({\blam}_0,r_0)}{\partial\blam\partial\blam^\intercal}= \frac{1}{n}\sum_{t=1}^{n}\frac{\partial^2\rho(\eps_t)}{\partial\blam\partial\blam^\intercal}\xrightarrow[n\to\infty]{a.s.} E\left[\frac{\partial^2\rho(\eps_t)}{\partial\blam\partial\blam^\intercal}\right]=\boldsymbol{H}(\bEta_0),$$
with $\boldsymbol{H}(\bEta)$ being defined in (\ref{eqn:matrices_def}). By combining Lemma 7.8 in \citet{Li13}, Assumptions (A4)-(A5) and technical arguments developed in \citet{Gor23}, one can show that there exists a constant $B>0$ such that
\begin{align*}
  \sup_{|r-r_0|<B/n}\left\|\frac{\partial\brho_n(\blam_0,r)}{\partial\blam} - \frac{\partial\brho_n({\blam}_0,r_0)}{\partial\blam}\right\|&=o_p(1),\\
  \sup_{\|\blam-\blam_0\|<B/\sqrt{n}}\;\sup_{|r-r_0|<B/n}\left\|\frac{\partial^2\brho_n({\blam},r)}{\partial\blam\partial\blam^\intercal} - \frac{\partial^2\brho_n({\blam}_0,r_0)}{\partial\blam\partial\blam^\intercal}\right\|&=o_p(1),
\end{align*}
hence we have
$$\sup_{|r-r_0|<B/n}\left\|\sqrt{n}(\hat{\blam}_n(r)-\blam_0)+\boldsymbol{H}(\bEta_0)^{-1}\frac{1}{\sqrt{n}}\frac{\partial\brho_n({\blam}_0,r_0)}{\partial\blam}\right\|=o_p(1)$$
and, in particular,
$$\left\|\sqrt{n}(\hat{\blam}_n(r_0)-\blam_0)+\boldsymbol{H}(\bEta_0)^{-1}\frac{1}{\sqrt{n}}\frac{\partial\brho_n({\blam}_0,r_0)}{\partial\blam}\right\|=o_p(1).$$
The proof is completed upon noting that
\begin{align*}
\sqrt{n} \sup_{|r-r_0|< B/n}&\left\|\hat{\blam}_n(r)-\hat{\blam}_n(r_0)\right\|\\
=  \sup_{|r-r_0|< B/n} &\left\|\sqrt{n}(\hat{\blam}_n(r)-\blam_0)+\boldsymbol{H}(\bEta_0)^{-1}\frac{1}{\sqrt{n}}\frac{\partial\brho_n({\blam}_0,r_0)}{\partial\blam}\right.\\
&+\left. \sqrt{n}(-\hat{\blam}_n(r_0)+\blam_0)-\boldsymbol{H}(\bEta_0)^{-1}\frac{1}{\sqrt{n}}\frac{\partial\brho_n({\blam}_0,r_0)}{\partial\blam}\right\|\\
\leq  \sup_{|r-r_0|< B/n} & \left\|\sqrt{n}(\hat{\blam}_n(r)-\blam_0)+\boldsymbol{H}(\bEta_0)^{-1}\frac{1}{\sqrt{n}}\frac{\partial\brho_n({\blam}_0,r_0)}{\partial\blam}\right\|\\
+ & \left\|\sqrt{n}(\hat{\blam}_n(r_0)-\blam_0)+\boldsymbol{H}(\bEta_0)^{-1}\frac{1}{\sqrt{n}}\frac{\partial\brho_n({\blam}_0,r_0)}{\partial\blam}\right\|=o_p(1).
\end{align*}
\par\noindent
\textbf{Point (iii)} From Point (ii), it holds that
$$\sqrt{n}(\hat{\blam}_n(r)-\blam_0)=\sqrt{n}(\hat{\blam}_n(r_0)-\blam_0)+o_p(1)= \boldsymbol{H}(\bEta_0)^{-1}\frac{1}{\sqrt{n}}\frac{\partial\brho_n({\blam}_0,r_0)}{\partial\blam}+o_p(1).$$
Note that the definition of $\bEta_0$ in (\ref{eqn:objective_pos}) implies that $\partial\rho({\blam}_0,r_0)/\partial\blam$ is a martingale difference sequence thereby the result follows by using the martingale central limit theorem:
$$\boldsymbol{H}(\bEta_0)^{-1}\frac{1}{\sqrt{n}}\frac{\partial\brho_n({\blam}_0,r_0)}{\partial\blam} \xrightarrow[n\to\infty]{d} N(\boldsymbol0,\boldsymbol{H}(\bEta_0)^{-1} \boldsymbol{J}(\bEta_0), \boldsymbol{H}(\bEta_0)^{-1}),$$
with $\boldsymbol{J}(\bEta)$ being defined in (\ref{eqn:matrices_def}).
\section{Analysis of the commodity time series}\label{SMsec:app}
\begin{figure}
\includegraphics[width=0.95\linewidth,keepaspectratio]{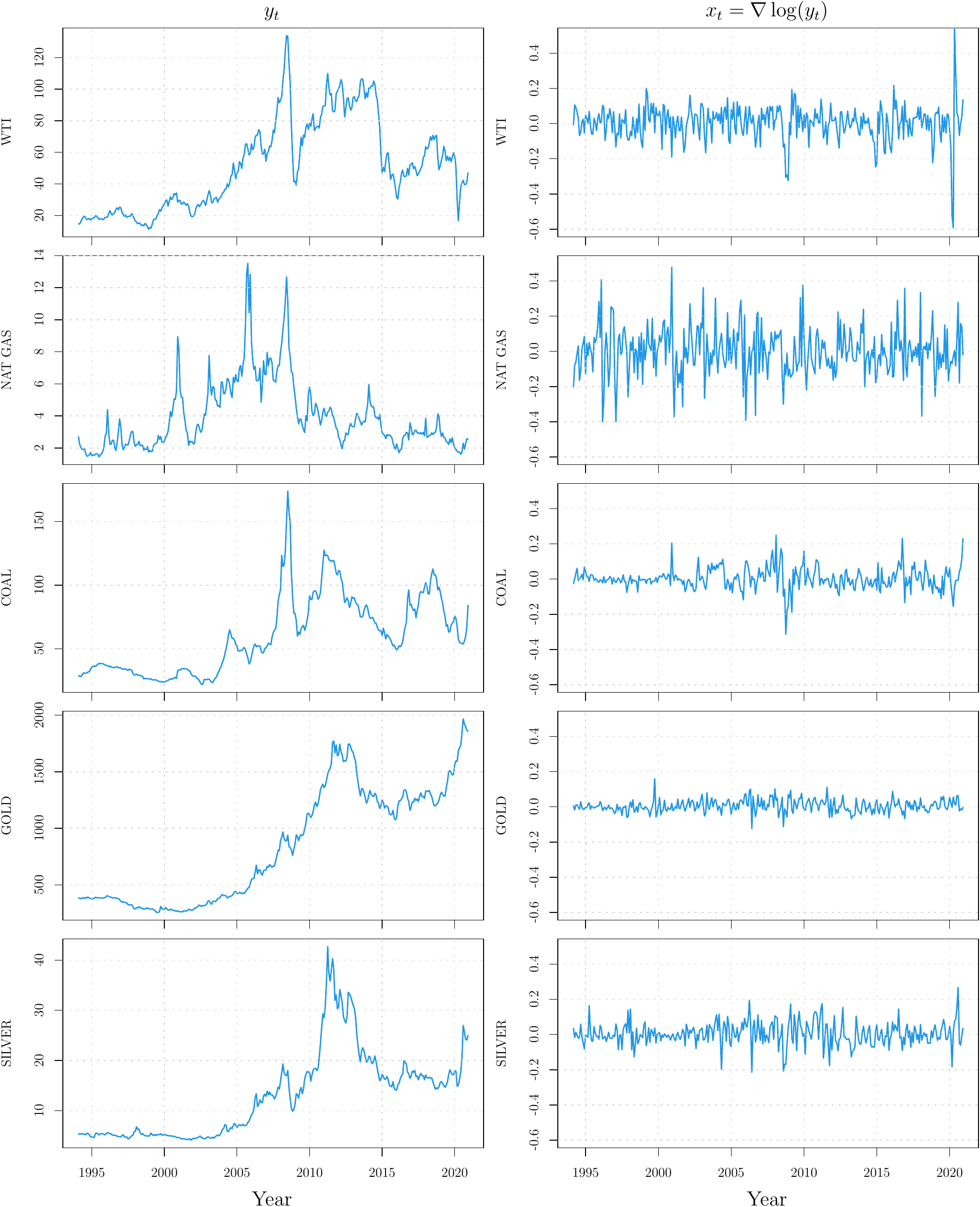}
\caption{Time plots of the five monthly commodities series February 1994 -- December 2020. Left column: raw commodities $y_t$. Right column: log-returns $x_t=\nabla \log(y_t)$.}\label{SMfig:1}
\end{figure}

\begin{figure}
  \centering
\includegraphics[width=0.7\linewidth,keepaspectratio]{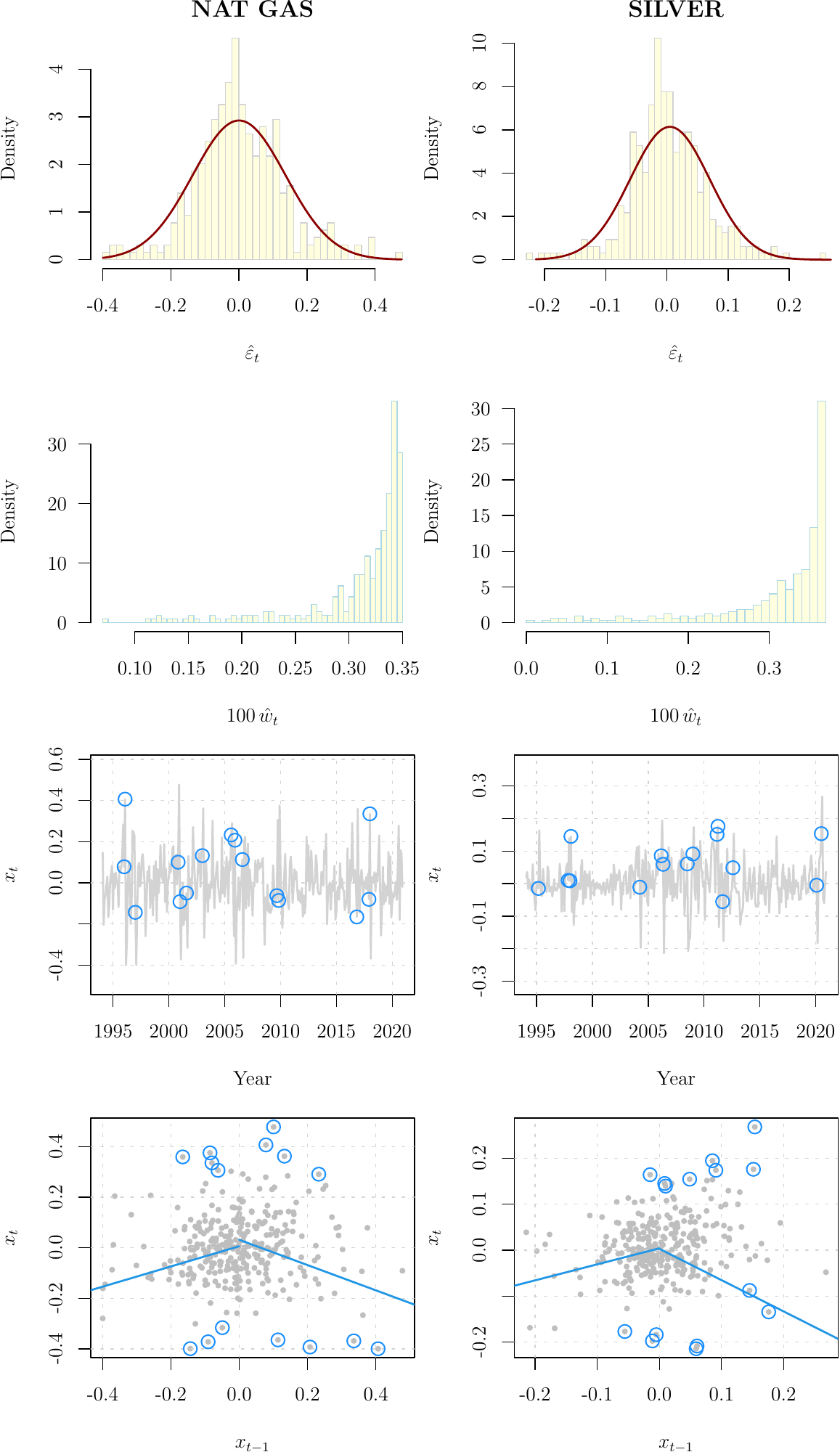}
\caption{Outlier analysis for the natural gas (NAT GAS), and silver (SILVER) commodity return series. Histogram of the residuals from the robust TARMA fit ($\hat \eps_t$, top row), robust weights ($100 \times \hat w_t$, second row), and the 15 most severe outliers shown by the circles in the time plot of the log-return series ($x_t$, third row) and in the corresponding lag plot of $x_{t}$ versus $x_{t-1}$ (bottom row).\label{SMfig:2}}
\end{figure}

In Figure~\ref{SMfig:1} we report the time plot of the monthly raw commodities $y_t$ (left column) and of their log-return $x_t=\nabla \log(y_t)$ (right column). In the latter, the plots use a common scale to highlight the different variability of the five series. The training set includes 324 monthly observations, from February 1994 to December 2021, and are obtained from the World Bank website \url{https://www.worldbank.org/en/research/commodity-markets}. The test set includes the 12 months of 2021 and is used to assess the performance of the models through the Mean Absolute Percentage Error, defined as follows:
\begin{equation}\label{SMeq:MAPE}
  \mathrm{MAPE}=\sum_{t=1}^{12}  \left|\frac{x_t - \hat{x_t}}{x_t}\right|\cdot 100,
\end{equation}
where $t$ ranges from January to December 2021.
In Figure~\ref{SMfig:2} we show the outlier analysis for natural gas and silver; see the main article for further details.
%

\bibliographystyle{abbrvnat}
\bibliography{robustTARMA_V5}

\end{document}